\documentclass[article]{jss}

\usepackage{orcidlink}
\usepackage{thumbpdf,lmodern}

\usepackage{framed}
\usepackage{float}
\usepackage{enumitem} 
\usepackage{float}

\usepackage{amsmath,amssymb}
\usepackage{booktabs}
\renewcommand{\vec}[1]{\mathbf{#1}}

\newcommand{\fct}[1]{\code{#1()}}
\newcommand{\mrm}[1]{\mathrm{#1}}
\newcommand{\e}{{\mathop{\mathrm{e}}}}

\author{Darren A. V. Scott~\orcidlink{0000-0001-8848-7770}\\AstraZeneca R\&D
   \And Sophia Axillus~\orcidlink{0000-0003-3427-3253}\\University of Gothenburg
   \And Alex Lewin~\orcidlink{0000-0003-0081-7582}\\LSHTM
   \AND Grant Izmirlian~\orcidlink{0000-0002-0481-631X}\\AstraZeneca R\&D}
\Plainauthor{Darren A. V. Scott, Sophia Axillus, Alex Lewin, and Grant Izmirlian}

\title{BayesFBHborrow: An \proglang{R} Package for Bayesian borrowing for time-to-event data from a flexible baseline hazard function}
\Plaintitle{BayesFBHborrow: An R Package for Bayesian borrowing for time-to-event data from a flexible baseline hazard function}
\Shorttitle{\textbf{BayesFBHborrow}: Flexible Baseline Hazard Borrowing in \proglang{R}}
\Abstract{Statistical methods that leverage external trial information to help accelerate drug development are becoming 
increasingly popular. Bayesian methods facilitate dynamic borrowing, where the similarity of the response guides how much 
information is used. We have proposed a semiparametric Bayesian borrowing model for time-to-event data, with smoothing 
priors that allows the baseline hazard to take any form via an ensemble average \citep{Scott2024}. By accurately modelling 
the baseline hazard, rather than approximating its form via fixed piecewise intervals, power is improved and bias of the 
estimated treatment effect reduced when the borrowing assumption of parameter exchangeability holds. A ``lump-and-smear'' 
borrowing prior makes the model robust to non-exchangeable historical data by increasing the sensitivity of borrowing to 
the presence of prior-data conflict, reducing the potential for type I error inflation.

We present \textbf{BayesFBHborrow}, an \proglang{R} package that implements our semiparametric Bayesian borrowing model with a
historical control. We demonstrate how to select the optimal borrowing hyperparameters. The model supports covariate-adjusted 
borrowing, which can reduce prior-data conflict and improve power when differences in outcomes are attributable to changes in 
the covariate distribution. As the treatment effect estimator is non-collapsible, the marginal hazard ratio can be estimated via 
Bayesian G-computation, while still permitting an adjusted analysis to account for control group drift. We illustrate the Bayesian 
flexible baseline hazard model on a simulated and real dataset with a marginal estimand, for both an unadjusted and adjusted analyses.

}

\Plainkeywords{Bayesian borrowing, flexible baseline hazard,   time-to-event, covariate adjustment, G-computation}

\Address{
  Darren A. V. Scott\\
  R\&D\\
  AstraZeneca\\
  1 Francis Crick Avenue, Cambridge, Cambridgeshire, UK, CB2 0AA\\
  E-mail: \email{darren.scott@astrazeneca.com}\\
  \\
  Alex Lewin\\
  Department of Medical Statistics\\
  London School of Hygiene and Tropical Medicine\\
  Keppel Street, London, UK, WC1E 7HT\\
  E-mail: \email{alex.lewin@lshtm.ac.uk}\\ 
    \\
 Sophia Axillus\\
    Department of Mathematical Sciences \\
    Chalmers University of Technology and University of Gothenburg \\
    S-412 96 Gothenburg, Sweden \\
    E-mail: \email{axillus@chalmers.se} \\
   \\
  Grant Izmirlian\\
  R\&D\\
  AstraZeneca\\
  101 Orchard Ridge Drive, Gaithersburg, Maryland 20874, USA\\
  E-mail: \email{grant.izmirlian@astrazeneca.com}
 }

\begin{document}


\section{Introduction} \label{sec:intro}


In early phase clinical trials, there has been a notable shift away from the traditional frequentist approach to Bayesian
methods. These methods are particularly beneficial in clinical research, as they allow for the incorporation of existing knowledge
with newly collected empirical data. This is relevant when conducting multiple trials for the same illness to discover effective
treatments or when evaluating established therapies. Bayesian borrowing, which involves integrating information from previous
trials into current research, can lead to improved efficiency through smaller and faster trials, increased statistical power, and
reduced patient allocation to less effective treatments.

The implementation of incorporating historical control data into the trial design is not without challenges. Researchers must
carefully consider the compatibility of historical data with new trial data to ensure that the modelling assumptions are
valid. This involves evaluating the similarity of patient populations, trial designs, and standard of care \cite{pocock}. Using
historical data that significantly diverge from current data can lead to issues such as an increased type I error rate and the
need for longer more costly trials to offset the influence of incompatible prior data or ``prior-data'' conflict. These
differences will be due to a data shift, either in the distribution of the prognostic covariate or the underlying relationship
between predictors and outcome.

The \pkg{BayesFBHborrow} \citep{BayesFBHborrow} package, available in \proglang{R} \citep{R}, enables borrowing from retrospective
data for time-to-event endpoints such as time-to-disease progression or time-to-death. These endpoints are the primary outcome in
a variety of therapeutic areas, including oncology \citep{oncology} and cardiovascular diseases \citep{cardiovascular}. The
borrowing is ``dynamic'', as the degree of historical data integration depends on the similarity of the data between the
historical and current data sets. We account for possible changes in the distribution of the prognostic covariates by
incorporating covariate adjustment to minimise prior-data conflict. By accurately capturing the true underlying hazard function in
our model, we improve the borrowing characteristics (power and type I error in the presence of prior-data conflict), compared with
approaches which constrain the hazard function (\cite{Scott2024}).

Bayesian borrowing is a relatively new field and its integration with time-to-event data is still developing. Packages such as
\pkg{RBesT} \citep{rbest} and \pkg{psborrow2} \citep{psborrow} represent the leading developments in the \proglang{R}
space. \pkg{RBesT} performs Bayesian borrowing by informing the prior for the new trial. First, the posterior predictive
distribution is sampled via Monte Carlo Markov Chain (MCMC) from a hierarchical model over the historical clinical trial summary
statistics. A mixture of densities, with estimated parameters that are conjugate with the likelihood of the new trial, is fitted
to the posterior predictive samples of the parameter of interest. By combining this with a vague conjugate prior and a suitable
weight, we have the robust meta-analytic predictive prior. This summarises the historical data, whilst accounting for the
uncertainty associated with a new parameter from the model and potential prior-data conflict. However, \pkg{RBesT} is not designed
to accommodate time-to-event data.

\pkg{psborrow2} incorporates external information by combining the likelihoods of the historical and the current data in one joint
model for either time-to-event or binary endpoints. A simple commensurate prior \citep{lewis}, which characterises the drift
between the current and historical parameter, controls the borrowing with either a Weibull or exponential likelihood. Thus, both
models place a constraint on the shape of the hazard function.

Various approaches have been proposed for Bayesian borrowing in the time-to-event setting. \cite{hobbs} posit a piecewise
exponential model (PEM) with a commensurate prior to control the level of borrowing. They assume predetermined fixed intervals and
independence across the associated baseline hazards. \cite{Han2017} use a PEM with fixed intervals and independent baseline
hazards over time, to borrow the control effect across multiple studies. They incorporate patient-level covariates to enhance the
efficiency of borrowing. \cite{lewis} use a Weibull likelihood and commensurate prior to borrow on the baseline hazard from the
one historical dataset. \cite{Bi2023} model the baseline hazard with an exponential distribution and use a Dirichlet process to
cluster the historical baseline hazards. This helps to discount the historical data in the presence of prior data conflict. All of
these methods impose a constraint on the shape of the baseline hazard over time, either as a horizontal line, a step function, or
a monotonic function.

In the \pkg{BayesFBHborrow} package we use a joint semiparametric hierarchical model to borrow information from a historical
time-to-event dataset with a smooth and flexible baseline hazard \citep{Scott2024}. We use an ensemble method, popular in
tree-based approaches such as random forests, to obtain a flexible baseline hazard function.
To smooth the shape of the baseline
hazard function, we introduce a dependency across the time points between the log baseline hazards via a nearest-neighbour
Gaussian Markov random field (NN-GMRF) prior \citep{Besag1995}, improving the accuracy of our estimation. ``Lump-and-smear" priors
can control the borrowing, making the borrowing robust to both violations of the assumptions of exchangeability and prior-data
conflict. The model can be used for either a marginal or conditional estimand, through either an adjusted or unadjusted model, as
baseline characteristics for the historical and current trial can be included and G-computation \citep{Robins1986} performed. This
allows the user to target a marginal hazard ratio, whilst the borrowing model is able to account for any drift explained by the
observed covariates, improving efficiency and reducing bias compared with simple marginal borrowing approaches.

Our proposed Bayesian borrowing model requires individual participant data (IPD), rather than summary data, as is the case with
some other recent approaches (\cite{Roychoudhury2020}, \cite{Bi2023}). Although this is clearly more difficult to obtain, there
are benefits in obtaining additional information, which is recognised by the FDA in its draft guidance
(\cite{FDA2026}). Whereas approaches which require a summary statistic constrain the shape of the baseline hazard, our model is
able to borrow information whilst remaining free to accurately capture the true shape of the underlying baseline hazard. The IPD
also allows the model to adjust covariates from either the current or historical trial (or both), which can help reduce posterior
uncertainty within our model. Again, this is not applicable for the summary-level data models.

\section{Model specification} \label{sec:models}
The \pkg{BayesFBHborrow} package performs Bayesian borrowing through an ensemble method popular in tree-based regressions, such as
random forests, to borrow external control information. For each iteration, we partition time and model the baseline hazard with a
step function. As we sample the parameters, the partitioned region and associated baseline hazard function are recursively altered
by changing the location of the split points or adding or removing split points (or both). The expected posterior baseline hazard
is a smoothed function obtained via an ensemble average. Because the baseline hazard function can be very flexible, we use our
prior structure to smooth its shape. This Section will guide the user through the theoretical basis of the sampler for the Bayesian Flexible Baseline Hazard (BFBH) model. For a more
detailed summary of the algorithm and the sampling methods, see \cite{Scott2024}.

\subsection{Likelihood}\label{sec:like}
In order to obtain a flexible baseline hazard we begin with a piecewise exponential likelihood for the current trial of the joint
model which consists of two elements, the hazard and the cumulative hazard
function at the time of the event. For the current trial consisting of $i = 1, ..., n$ patients with the time-to-event outcome
represented by $y_i$ and event indicator $\nu_i$. We divide the time axis into $J + 1$ partitions with split points $0 = s_0 < s_1
< ... < s_{J+1}$, and assume a constant baseline hazard $h_j(y_i)$ per partition for $y_i \in I_j = (s_{j-1}, s_j)$.


The vector of stratification variables or prognostic covariates ($p \times 1$) is defined by $\mathbf{x}'_i = (x_{i1}, ...,
	x_{ip})$ for subject $i$ with corresponding regression coefficients $\boldsymbol{\beta} = (\beta_1, ..., \beta_p)$ and treatment
covariate $Z_i$ and associated effect $\varphi$. Let the interval specific baseline hazards be defined as $\boldsymbol{\lambda} =
	(\lambda_1, ..., \lambda_{J+1})$, the likelihood of the current data is
\begin{align}
	\begin{aligned}\label{eq:likelihood}
		\mathcal{L}( \boldsymbol{D}| \boldsymbol{\beta}, \boldsymbol{\lambda}, \varphi, \mathbf{s}, J ) = & \prod_{i=1}^{n}\prod_{j=1}^{J+1}
		\big(\lambda_j\exp(\mathbf{x}_i'\boldsymbol{\beta}+ Z_i\varphi)\big)^{\delta_{ij}\nu_i}\exp\Big(-\delta_{ij}\big(
                \lambda_j(y_i - s_{j-1})\\
            & + \sum_{g=1}^{j-1}\lambda_g(s_g-s_{g-1})\big) \exp(\mathbf{x}_i'\boldsymbol{\beta} +Z_i\varphi) \Big),
	\end{aligned}
\end{align}
where $\boldsymbol{D} = (n, \boldsymbol{y}, Z, \mathbf{X}, \boldsymbol{\nu})$ denotes the observed data with $\nu_i = 1$ if the
$i$th subject failed (event happens) and 0 otherwise, and the variable $\delta_{ij}$ is 1 if the subject $i$ was censored or
failed in interval $j$, and 0 otherwise. 

The likelihood for the historical data will have the same structure and split point parameters, defined as $\mathcal{L}
	(\boldsymbol{D}_0| \boldsymbol{\beta}_0, \boldsymbol{\lambda}_0, \mathbf{s}, J )$ with $\boldsymbol{D}_0 = (n_0, \boldsymbol{y}_0,
	\mathbf{X}_0, \boldsymbol{\nu}_0)$.

\subsection{Priors}
The total number of split points $0= s_0 < s_1, ..., s_{J+1}$ with $s_0$ and $s_{J+1}$ fixed, has a right-truncated Poisson
distribution with hyperparameter $\phi$ and right truncation limit  $J_\text{max}$
\begin{equation}\label{eq:phi}
	J \sim \text{Poisson}_\text{t}(\phi) \qquad 0 \leq J \leq J_{\text{max}}.
\end{equation}
The truncation shrinks the mean number of split points and to a greater extent the variance below
$\phi$, whilst controlling the flexibility of the baseline hazard 
and reducing the computational burden of the sampler. The distribution of the split point locations, is the even-numbered order statistic of $2J + 1$ uniformly distributed points \citep{green}.

To smooth the posterior baseline hazard we introduce a dependency of the historical log baseline hazards on their nearest
neighbour through a GMRF prior
\begin{equation}
	\log (\boldsymbol{\lambda}_0)|\mu, \sigma^2_\lambda \sim \mathcal{N}_{J+1}(\mu \boldsymbol{1}, \sigma_{\lambda}^2\Sigma_s(c_\lambda)), \label{eq:gaussianmarkov}
\end{equation}
with means $\mu$ and the covariance matrix $\sigma_{\lambda}^2\Sigma_s(c_\lambda)$. The overall variability of the historical log baseline
hazard is defined by $\sigma_{\lambda}^2$ and the smoothness is controlled by the hyperparameter $c_\lambda$ $\in (0,1)$ within
$\Sigma_s$ (denoted as a function of $c_\lambda$). Additionally, these parameters have the following hyperpriors
\begin{align}
	\sigma_\lambda^2 & \sim \text{Inv-Gamma}(a_\sigma, b_\sigma), \label{eq:sigma_lambda} \\
	\pi(\mu)         & \propto 1. \label{eq:mu}
\end{align}

The priors for the treatment effect, regression coefficients corresponding to the baseline characteristics for the current and
historical data are
\begin{equation}\label{eq:betas}
	\varphi \sim \mathcal{N}(0,\sigma_\beta^2), \qquad  \boldsymbol{\beta} \sim \mathcal{N}(0,\sigma_\beta^2) , \qquad \boldsymbol{\beta}_0 \sim \mathcal{N}(0,\sigma_{\beta_0}^2).
\end{equation}
with $\sigma_{\beta}^2$ and $\sigma_{\beta_0}^2$ sufficiently large for a vague priors.

To account for a discrepancy between historical and current data, we use a commensurate prior to borrow on the historical log
baseline hazard \citep{hobbs}
\begin{equation}
	\label{eq:lambda_clam}
	\log (\lambda_j)|\tau_j \sim \mathcal{N}
	(\log(\lambda_{0j}), \tau_j),
\end{equation}
where $\tau_j$ is a commensurability (variance) parameter.

\subsection{Borrowing method}
It is the prior on $\tau_j$ that controls the borrowing, and this has two different prior parameterisation options, either the
borrowing is determined per split point or across all the split points (where $\tau_j = \tau)$. Both of these approaches are
modelled by a mixture of inverse gamma distributions and mixture weight $p_0$.
\begin{align}
	\tau_j^{(\text{mix})} & \sim p_0 \text{Inv-Gamma}(a_\tau, b_\tau) + (1-p_0)\text{Inv-Gamma}(c_\tau, d_\tau), \label{eq:mix}  \\
	\tau^{(\text{all})}   & \sim p_0 \text{Inv-Gamma}(a_\tau, b_\tau) + (1-p_0)\text{Inv-Gamma}(c_\tau, d_\tau).  \label{eq:all}
\end{align}

There is also the option of a simple prior (which is nested in prior (\ref{eq:mix})) with $p_0 = 1$
\begin{equation}\label{eq:uni}
	\tau_j^{(\text{uni})} \sim \text{Inv-Gamma}(a_\tau, b_\tau).
\end{equation}
The inverse gamma prior of (\ref{eq:uni}) imposes a simple assumption of exchangeability between the baseline hazards for the
interval $I_j$. The lump-and-smear priors (\ref{eq:mix}-\ref{eq:all}) allow for the possibility of large differences between the
log baseline hazard of the historical and current data, which robustifies the prior against prior-data conflict.

\subsection{Choice of hyperparameters} \label{sec:hyper_choice}
As the selection of the hyperparameters which control the borrowing are scale dependent, by default the package will standardise
time in the current data so that there is an average of one event per unit of time. This standardisation constant will
also be applied to the historical data simplifying the model specification, so that the principled approach to hyperparameter selection for dynamic borrowing outlined in \cite{Scott2024} can be followed. The MCMC posterior samples and associated plots from
the package are returned in the original unit of time.

\subsubsection{Smoothing parameters}
When choosing the smoothing parameters of $\phi$, $J_\text{max}$ and $c_\lambda$, our aim is to exploit the trade-off between
flexibility and variability. Although the baseline hazard can take more unusual shapes when $\phi$ is large, this is typically
unrealistic. \cite{Scott2024} explore combinations of smoothing parameters for various shaped baseline hazard functions. They find
that the best hyperparameter choice for their simulated historical datasets based on the fit of the model, regardless of the
underlying hazard, is $\phi = 3$ and $J_\text{max}$ = 5. 

The optimal choice of $c_\lambda \in (0,1)$ can be made with an understanding of the shape of the baseline hazard. A regular shape
hazard function requires a large $c_\lambda$, while an irregular shape requires a small value. As the true shape of the hazard
function is unknown, the expected shape can be estimated by running the model, without the borrowing structure, on the external control group data.

\subsubsection{Borrowing parameters}
\cite{Scott2024} explore the impact of these hyperparameters on dynamic borrowing, through borrowing profiles which define how
quickly historical information is discounted for discrepancies between the log-baseline hazard. They demonstrate that specific choices of
the mixture prior hyperparameters induce favourable borrowing characteristics, close to full borrowing for tolerated differences
between the log baseline hazard, whilst information is fully discounted when prior-data conflict is detected by the model.  A simple approach to selecting appropriate values is proposed that is agnostic to the units of time the events are measured in and
enables the user to specify their tolerance for differences between the log-baseline hazard. The approach for the mix prior
parameterisation is;
\begin{itemize}
	\item Standardise the current and historical baseline hazard so that there is an average of one event per unit of time
	      in the current data  (performed by the package). This means that the control hazard is approximately 1.
	\item Set $a_\tau= c_\tau=1$ and $b_\tau = 0.001$ and $d_\tau=5$.
	\item Define a limit of tolerable difference between the standardised log baseline hazards $\xi$. This is the point at which
	      the posterior weight will drop below 0.5.
	\item Calculate the prior weight
	      \begin{equation}\label{eq:tip_p}
		      p_{0} = \left( 1 +  \frac{0.001}{d_\tau}  \left( \frac{\xi^2 + 0.002  }{\xi^2 + 2  d_\tau } \right)^{-\frac{3}{2}}
		      \right)^{-1}.
	      \end{equation}
\end{itemize}

The posterior weight (or borrowing profile) $q_0$, can be plotted as a function of the potential difference between the log
hazards for the choice of hyperparameters
\begin{equation}\label{eq:bb_profile}
	q_0 = \left( 1 +  \frac{1-p_0}{p_0}\frac{d_\tau}{b_\tau}  \left( \frac{(\log(\lambda_j)-\log(\lambda_{0j}))^2 + 2 b_\tau }{
		(\log(\lambda_j)-\log(\lambda_{0j}))^2 + 2  d_\tau } \right)^{-\frac{3}{2}}\right)^{-1}.
\end{equation}
Alternatively, a prior weight can be defined first, the limit of tolerable difference between the log baseline hazards calculated
in (\ref{eq:bb_profile}) and the borrowing profile plotted.

If $\tau^{(\text{all})}$ (\ref{eq:all}) is used in the model, then the limit of tolerable difference $\xi$ in
(\ref{eq:tip_p}) will be replaced by the limit of the sum of squared error. This will be harder to define, so the user may wish to
plot various borrowing profiles for different choices of $p_0$ to decide on a suitable value.

If the simple univariate inverse gamma prior (\ref{eq:uni}) is selected, then $b_\tau$ should be set to a small value (typically
0.001) to encourage borrowing. However, this prior imposes a simpler assumption of exchangeability between the baseline hazards for
the interval $I_j$, and the lump-and-smear mixture priors, which robustifies the model against prior-data conflict, are
preferable. 

\subsection{Design matrix for an adjusted analysis}\label{sec:DM}
If covariates are to be included in the model through $\mathbf{X}_0$ and $\mathbf{X}$ for an adjusted analysis, care must be taken
to ensure that all historical information is utilised and the current and historical baseline hazard remain concordant. However, any type of design matrix can be used, such as reference-based or sum-to-zero, as the baseline hazard is common to all subjects the parameterisation for the concordant design matrices does not impact the borrowing.

\subsection{MCMC sampler and posterior inference}
To sample from the posterior distribution, we use a Metropolis-Hastings \citep{MH} within Gibbs \citep{gibbs} sampler. The sampler
can be initialised with default settings, but adjustments will need to be made to the proposal variance to ensure that the MCMC is
sampling the joint posterior sufficiently. We provide a brief summary of the samplers to help the user make the changes correctly
and outline the arguments that control the output for the posterior inference.

For fixed $J$ the unknown parameters in the joint model are
\begin{equation}
	\boldsymbol{\theta}(J) =(\varphi, \boldsymbol{\beta}, \boldsymbol{\lambda}, \boldsymbol{\beta}_0,  \boldsymbol{\lambda}_0, \boldsymbol{s}, \boldsymbol{\tau},
	\sigma^2_\lambda, \mu),
\end{equation}
the components of $\boldsymbol{\theta}(J)$ are updated by either exploiting conjugacies in the full conditionals or via
Metropolis-Hastings steps.

The posterior treatment effect $\varphi$, current $\boldsymbol{\beta}$ and historical $\boldsymbol{\beta}_0$ regression
coefficients posterior are sampled with a Metropolis-Hastings step. As the gradient and observed curvature of the log-posterior
are available in closed form, we employ a Newton-Raphson, locally Gaussian proposal for both the historical and current regression
coefficients. At each iteration, we form a multivariate, curvature-adjusted normal approximation around the current parameter
value and propose from this distribution, using a single scalar step-size to scale the stochastic component for each coefficient
vector, $(\varphi, \boldsymbol{\beta})$, and $\boldsymbol{\beta}_0$. These tuning parameters, \code{cprop_beta} and
\code{cprop_beta_0} respectively, need to be varied in shorter runs so that the probability of accepting each multivariate
proposal is approximately $40\%$, so that the posterior distribution is explored fully.

For the historical baseline hazard we use a Metropolis-Hastings where the proposal is from a conjugate posterior, after an
independent vague gamma prior is combined with the historical likelihood. For the current baseline hazard, the vague gamma prior
is combined with both the historical likelihood discounted by $\text{Gamma}(\text{shape} = a_\lambda, \text{rate} = b_\lambda)$
and the current likelihood
\begin{equation}
	\pi_{(\text{prop})}(\lambda_j|\boldsymbol{D}, \boldsymbol{D}_0,\alpha) \propto
	\mathcal{L}(\boldsymbol{\lambda},\boldsymbol{\beta}, \boldsymbol{s}|\boldsymbol{D})
	\mathcal{L}(\boldsymbol{\lambda}_0,\boldsymbol{\beta}_0, \boldsymbol{s}|\boldsymbol{D}_0)^\alpha \pi(\lambda_j),\label{eq:alpha}
\end{equation}
where $0 \leq \alpha \leq 1$. The power parameter \code{alpha} is set by default at 0.4 to increase the variability of the
proposal by discounting historical information. However, you may need to increase both $a_\lambda$ and $b_\lambda$ and (or) change
$\alpha$ to ensure an appropriate acceptance ratio. As the data is standardised in the sampler so that the baseline hazard
is approximately 1, $a_\lambda$ and $b_\lambda$ can remain equal.

Our model treats the split points, both in terms of their location and the total number, as random. The total number of splits $J$
are sampled via a reversible-jump MCMC \citep{green} (RJMCMC) which extends or reduces the number of split points by one. This
either adds or deletes a split point, whilst adjusting the other parameters in the model.  The probability of a birth move (and
the respective death move) is set to 0.5 by default, but this can be changed by the user.

To obtain the smoothed posterior baseline hazard, a discrete grid of dimensions $n_{\text{iter}} \times n_{\text{grid}}$ is
constructed, where $n_{\text{iter}}$ is the number of iterations of the sampler and $n_{\text{grid}}$ is the total number of
equally spaced partitions of the time interval from 0 to $\mbox{max}(y_i|\nu_i = 1)$. For each iteration, the baseline hazard
associated with the corresponding time interval is saved in the grid. The smoothed expected posterior baseline hazard is an
ensemble average
\begin{equation}
	\label{eq:smooth_bh}
	\lambda(t) = \frac{1}{n_{\text{iter}}}\sum^{n_{\text{iter}}}_{\nu=1} \lambda_{j}(\nu),
\end{equation}
where $\lambda_j(\nu)$ is the interval containing $t$ in the MCMC iteration $\nu$ and $n_{\text{iter}}$ is the number of
iterations. The smoothness of these parameters is controlled by $\phi$, $J_\text{max}$ (\ref{eq:phi}) and $c_\lambda$
(\ref{eq:lambda_clam}).



The package also provides plots of the smoothed expected posterior predictive hazard function
\begin{equation}
	\label{eq:smooth_pred_bh}
	\lambda(t|\vec{x}_i)_i = \frac{1}{n_{\text{iter}}}\sum^{n_{\text{iter}}}_{\nu=1} \lambda_{j}(\nu)\exp \left\{\mathbf{x}'_i\boldsymbol{\beta}(\nu) + z_i\varphi(\nu) \right\},
\end{equation}
and expected posterior predictive survival function
\begin{equation}
	\label{eq:smooth_pred_sf}
	S(t|\vec{x}_i)_i = \frac{1}{n_{\text{iter}}}\sum^{n_{\text{iter}}}_{\nu=1} \exp\left\{-\int_0^t
	\lambda_{j}(\nu)\exp \left\{\mathbf{x}'_i\boldsymbol{\beta}(\nu) + z_i\varphi(\nu) \right\}d\nu
	\right\},
\end{equation}
where $\boldsymbol{\beta}(\nu)$ and $\varphi(\nu)$ are the posterior samples of the regression coefficients and the treatment
coefficient for an individual with a covariate set $\mathbf{x}_i$ in the treatment group $z_i$. These are plotted along with the
associated credible intervals.

\section{Estimand and adjusted models}\label{sec:est and adj}

In Section~\ref{sec:PF}, we demonstrate how to use the package under different population-level summaries, a key element of the estimand framework \citep{Gogtay2021}. We adopt the terminology
from \citep{Daniel2021}, where we use conditional/marginal to refer to the estimand (or parameter that defines our target
treatment effect) and adjusted/unadjusted to refer to the analysis model. In our setting,  as the estimand of interest in our time-to-event setting is non-collapsible, we cannot simply use an adjusted model (which is typically more robust to prior-data conflict) if we wish to target a marginal estimand. In the first example, rather than omitting covariates from the model to estimate our marginal treatment effect, we account for any potential drift of the historical control by first performing a covariate adjusted analysis and then marginalising over the adjusted
probability measure using an approach called standardisation or Bayesian G-computation \citep{Keil2018}.  This approach sits naturally within the Bayesian framework, where both marginalisation and inference are simple steps that are performed with the MCMC sampled posterior
distribution. In the second example, our  focus is on a conditional estimand.  After a discussion on covariate adjustment and the implications in our model, we provide a short summary of
G-computation for our choice of estimand with our time-to-event data, before illustrating the application of our model.

\subsection{Covariate adjustment}
In many of the Bayesian dynamic methods that borrow control information, the focus is on borrowing with a marginal sufficient
statistic. However, the control outcome is likely to be affected by variables, for example, in the oncoloy setting the tumour
stage is predictive of survival. If the distribution of these key prognostic covariates varies across the current and control
trial, either due to differences in the study's target population or ``drift'' as the standard of care improves over time or chance,
then the possibility of prior-data conflict increases. By including prognostic covariates in an adjusted model, we can account for
this ``imbalance'', reducing the risk of prior-data conflict.

When we adjust for covariates in our model through ($\mathbf{X}, \mathbf{X}_0$), the assumption of exchangeability of the current
and historical log baseline hazard in (\ref{eq:lambda_clam}) is now conditional on the covariates. For borrowing to be meaningful,
the true log-hazard function of the covariates must be the same in both the historical and control datasets. 
We recommend
examining the current and historical covariate effect posterior summaries to assess how reasonable this assumption is in your
application. The assumption that both datasets are from the same source population; under similar measurement, inclusion criteria,
clinical practice, and follow-up processes is even more important \citep{pocock} when working with an adjusted model.


Unfortunately, the log link function in our joint hazard ratio model leads to a non-collapsible treatment estimand, if the
treatment is effective (all effect measures are collapsible for null treatment effects) (\cite{Daniel2021}). The treatment
estimand from a model with covariate adjustment (even with an omission of any treatment covariate interaction) does not have a marginal
interpretation, the conditioning changes the nature of the treatment effect we are estimating. In the following subsection we describe a
procedure which allows us to adjust for covariates and increase precision, whilst still targeting a marginal estimand. This
highlights another aspect of time-to-event modelling, the effect of individual heterogeneity or frailty and drop-out on the
estimated treatment effect (in our case the hazard ratio). This phenomenon has also been widely discussed in the literature,
particularly in the context of a causal interpretation (\cite{Fay2024}, \cite{Hernn2010}, \cite{Aalen2015}).

\subsection{G-computation and marginal estimands from adjusted models}
\label{sec:G_comp}

The marginalisation procedure ensures we exploit both covariate adjustment and available external data borrowed in a robust
fashion, to optimise the power for testing a marginal treatment estimand. Intuitively, as we are interested in population-level
effects, we work with the survival distribution rather than the hazard function (which is conditional on surviving up to the
argument of the function, time $t$) directly. The marginalised survival function for treatment $Z$, from expressing the survival
probability in terms of the cumulative hazard function is
\begin{align}
	\label{eq:marg_surv}
	S(t|Z=z, \boldsymbol{\theta}) & =\mathbb{E}[S(t|Z=z,\vec{X},\boldsymbol{\theta})] \nonumber                                       \\
	                              & = \int_{\vec{X}} \exp(-H_0(t))^{\exp (\vec{x}'\boldsymbol{\beta} + z\varphi )}p(\vec{X})d\vec{X},
\end{align}
where the expectation is over the covariate distribution and $H_0(t)$ is the cumulative baseline hazard function given the
baseline hazards $\lambda$, and time partition parameters $\vec{s}$ and $J$,
\begin{equation*}
	\label{eq:cum_hazard}
	H_0(t|\boldsymbol{\theta}) =
	\prod_{j=1}^{J+1} \delta_{ij}\big( \lambda_j(t - s_{j-1}) \\
	+ \sum_{g=1}^{j-1}\lambda_g(s_g-s_{g-1})\big)
\end{equation*}
with $\delta_{ij} =1$ when $t$ is in the $j$th interval and 0 otherwise.

Rather than performing a full Bayesian analysis, in which a joint prior distribution is specified for the covariates (discrete,
continuous, or both) (\cite{Keil2018}), we adopt an approximation procedure based on the Bayesian bootstrap
(\cite{Willard2024}). For each unique covariate pattern $\vec{x}_i= \vec{x}_{k_i}$, $k=1, ...,K$, we specify a Dirichlet prior on
the pattern weights
\begin{align*}
	\boldsymbol{\pi} \sim                & \text{Dirichlet}(\alpha_1,...,\alpha_K),            \\
	k_i \sim                             & \text{categorical}(\boldsymbol{\pi}),               \\
	\boldsymbol{\pi}|\boldsymbol{k} \sim & \text{Dirichlet}(\alpha_1 + k_1,...,\alpha_K + k_K)
\end{align*}
where $k = (k_1, ..., k_K)$ are the counts of observations in each unique covariate pattern. As is standard practise
\cite{Rubin1981}, we set $\alpha_i=0$ for all $i=1,...,K$. The conjugate posterior weights $\boldsymbol{\pi}|\boldsymbol{k}$
define a random discrete distribution over the observed covariate patterns and are used to approximate the integral over
$p(\vec{X})$ via a Bayesian bootstrap. The marginalised survival distribution can then be transformed to obtain posterior samples
of the log hazard ratio $\gamma(\boldsymbol{\theta})$
\begin{equation*}
	\gamma(\theta)_m = \log(-\log(S(t|Z=1,\boldsymbol{\theta}_m))- \log(-\log(S(t|Z=0,\boldsymbol{\theta}_m)).
\end{equation*}
The full procedure in the package is described in the Appendix \ref{app:g-computation}. Our routine returns the marginalised
hazard for each treatment group over a grid of time. This allows the user to specify a time point, plot the posterior expectation
and variance over time, or report a single measure across time.

The interpretation of hazards and hazard ratios for comparing treatment and exposure groups is complicated by the possible effects
of unmeasured frailty on event times \citep{Fay2024}.  Under the model structure in (\ref{eq:likelihood}), as we assume hazards
conditional on the split point and covariates
\begin{equation}
	\label{eq:model_assum}
	\lambda_j(t)\exp(\boldsymbol{x}_i'\boldsymbol{\beta} + Z_i \varphi)
\end{equation}
are correctly specified, the model implied by the G-computation step will, in general, be misspecified. This is because the
composition of the risk sets under each arm at any given time generally differs from that at the start of the trial, because under
a non-null treatment effect, patients with higher hazards tend to experience events earlier. This dynamic selection alters the
balance of covariates and unmeasured risk factors in the risk set over time, despite being exchangeable at baseline due to
randomisation. Consequently, the true marginal hazard ratio typically varies with time, an effect that is not captured when
fitting a marginal proportional hazards model that assumes a constant hazard ratio. This emphasises the importance of interpreting
a hazard as, the instantaneous event rate in the group of survivors at each time, rather than a common hazard applicable to each
individual in the study.

This limitation reflects the behaviour of marginal proportional hazards models when an underlying conditional relationship is
omitted, which is well documented in the literature (\cite{Hernn2010}, \cite{Stensrud2017}).  In randomised controlled trials, the
marginal estimand is often targeted, and a simplifying assumption of time-constant treatment effect (a constant marginal hazard
ratio) is made, despite the likely presence of frailty. In this case, our covariate-adjusted estimator from the g-computation
procedure will have a very similar expectation to the unadjusted estimator but with increased precision.

Under a non-null treatment effect, the effect of time varying risk sets in time-to-event data due to unmeasured frailty will
affect the marginal posterior hazard ratio of our model, even without additional prognostic covariate adjustment. This is because
we obtain a time-varying hazard, by assuming a constant hazard (and constant hazard ratio) for each time partition, and then
average over unmeasured frailty and time partitions rather than conditioning on them. The balance of the treated and control group
will change dynamically as higher-risk or ``sicker'' patients fail earlier, leading to a slight attenuation toward the null of the
marginal hazard ratio, which is no longer constant over time.

\section{Package Functionality through Examples}
\label{sec:PF}
The design philosophy is based upon the premise that this tool is intended for the analysis of survival data. Accordingly, the
canonical \proglang{R} methods for providing a summary, extracting coefficients, and generating plots are primarily aimed at
presenting the results of survival analysis, with a secondary focus on producing diagnostics for the MCMC sampler. In this Section 
we explain the essentials of the package functionality, e.g. calling sequence, meaning of arguments, and most importantly, the
connection between the methodology presented in the previous sections and consequences of various options provided by the package,
as well as proper interpretation of the results. This will be presented within the context of a series of related examples using
data generated via an included simulation function, \code{genBFBHBdat}, to conclude with analysis of the German Breast Cancer
Study (GBCS) of Tamoxifen which is publicly available in the \proglang{R} package, \pkg{condSURV}. 

\subsection{Examples using simulated data}
The simulated current trial and historical control dataset are obtained via the included data generation function \fct{genBFBHBdat}. We begin by
specifying the sample sizes and true values of the parameters, where the variable names are assigned to their
corresponding argument names. First,  \code{n_cc_1} and \code{n_cc_0} are the sizes of the current trial treated and control arms,
and \code{n_hst} is the size of the historical controls data set. 
\begin{Schunk}
\begin{Sinput}
R> n_cc_1 <- 200
R> n_cc_0 <- 100
R> n_hst <- 100
\end{Sinput}
\end{Schunk}
The data generating mechanism is a proportional hazards Weibull model. The true values of the shape parameter and the effect of treatment in the current 
trial are \code{shape} and \code{B_trt}.
\begin{Schunk}
\begin{Sinput}
R> shape <- 2
R> B_trt <- log(0.55) 
 \end{Sinput}
\end{Schunk}
In this example, we will have 3 pretreatment covariates consisting of two continuous variables and a factor variable. This is specified via the coefficient 
vector arguments, \code{B_x_cc}, \code{B_x_hst}, and the argument \code{X_fact_levs} in the following way. The argument \code{X_fact_levs} gives the number
(its length) and levels (its components) of factor variables, so for one factor variable of three levels, we set \code{X_fact_levs=3}. This will require two non-referent effects in both of the coefficient vectors. Continuous covariates are specified by filling the coefficient vectors out further on the front end. Notice below that the coefficient vectors are of length 4 corresponding to a single factor variable with two non-referent levels and two continuous covariates. Here we use equal pretreatment coefficient vectors for the current trial and historical trial, but this need not be the case in general. Lastly, we specify the intercepts, \code{int_cc} and \code{int_hst}, in the current trial and historical controls linear predictors. They are also taken here to be equal, but this need not be the case. 
\begin{Schunk}
\begin{Sinput}
R> X_fact_levs <- 3
R> B_x_cc <- B_x_hst <- c(-0.3,0.5,0.25,-0.5)
R> int_cc <- int_hst <- -log(3)
\end{Sinput}
\end{Schunk}

We call our data generation function, which returns a list of \code{data.frame}s, named \code{DAT_cc} and \code{DAT_hst}, and assign them to the corresponding objects. 
\begin{Schunk}
\begin{Sinput}
R> dat_lst <- genBFBHBdat(n_cc_1=n_cc_1, n_cc_0=n_cc_0, n_hst=n_hst,
+                         B_trt=B_trt,
+                         B_x_cc=B_x_cc, 
+                         B_x_hst=B_x_hst,
+                         int_cc=int_cc, int_hst=int_hst, shape=shape,
+                         t_er=0.5, t_fin=1.5, X_fact_levs=X_fact_levs)
R> DAT_cc <- dat_lst$DAT_cc
R> DAT_hst <- dat_lst$DAT_hst
\end{Sinput}
\end{Schunk}

\subsubsection{Selection of Hyperparameters}
We outline two approaches to model hyperparameter selection. In our first set of examples based upon simulated data, rather than
following Section \ref{sec:hyper_choice} in full to select our borrowing hyperparameters, we outline an abbreviated approach according to an effective sample size (ESS). The borrowing is controlled by the
hyperparameters in the $\tau_j$ hyperprior. To robustify the model, a lump-and-smear mixture of inverse gamma distributions
(\ref{eq:mix}) is selected via \code{model_choice = "mix"}. We fix \code{a_tau} = \code{c_tau} = 1, define the lump to encourage
borrowing with \code{b_tau} = 0.001 and choose the vague smear component with \code{d_tau} = 5 as in the principled approach
outlined in Section \ref{sec:hyper_choice}. Our choice of borrowing hyperparameters ensures a sufficiently diffuse density for the
vague component for the standardised data, so that a sample of the commensurate parameter is likely to fully discount the external
information. Our prior weight of $p_0 = 0.8$ can therefore be interpreted to have an approximate effective sample size of $80\%$
of the historical sample size. Solving (\ref{eq:tip_p}) for our choice of $p_0$, means that our commensurate mixture prior can be
interpreted as defining a limit of tolerable difference between current and historical log hazards ($\xi$) on a standardised scale
of 0.29 (the call function is below) or a hazard ratio of 1.35. This is the threshold for data conflict, as the posterior weight
associated with the component which encourages borrowing drops below 0.5.
\begin{Schunk}
\begin{Sinput}
R> xifinder(b_tau = 0.001, d_tau = 5, p_0 = 0.8)
\end{Sinput}
\begin{Soutput}
[1] 0.2913806
\end{Soutput}
\end{Schunk}

The smoothness of the baseline hazard is controlled by the truncated Poisson prior on the split points and the NN-GMRF prior hyperparameter $c_\lambda$. To manage the trade off between flexibility and variability, we follow the example
of \cite{Scott2024} who demonstrate the importance of restricting the split points to prevent the model from over fitting the
posterior baseline hazard to the data, with \code{phi}=3 and \code{Jmax}=5. This constrains the flexibility of the smoothed
posterior baseline hazard,the wider time intervals within the PEM ensure that there is more data to estimate the baseline hazard
at later time-to-events, where there are fewer events, reducing the overall variability of the baseline hazard estimate. As we
know the underlying baseline hazard is monotonic, we choose a \code{clam_smooth} of 0.8.  The default setting of a vague prior choice of (1,1) for \code{a_sigma} and \code{b_sigma} (\ref{eq:sigma_lambda}) controls the overall variability of the
historical baseline. Finally, the variance of the vague prior on the treatment parameter is set by \code{beta_prior}=$10^2$.

\begin{Schunk}
\begin{Sinput}
R> hyperparameters_sim < list(beta_prior = 10^2,
+                             beta_0_prior = 10^2,
+                             a_tau = 1,
+                             b_tau = 0.001,
+                             c_tau = 1,
+                             d_tau = 5,
+                             p_0 = 0.8,
+                             a_sigma = 1,
+                             b_sigma = 1,
+                             clam_smooth = 0.8,
+                             phi = 3,
+                             Jmax = 5)
\end{Sinput}
\end{Schunk}
The MCMC sampler will select the initial parameter values. The historical baseline hazard is sampled from a gamma distribution
with shape and scale equal to the number of events and the total exposure time within the historical data. The estimated mean and
standard deviation of the log transformed data is used for the initial values of $\mu$ and $\sigma_{\lambda}^2$ in the NN-GMRF
prior (\ref{eq:gaussianmarkov}). The commensurate parameters, $\tau_j$ are sampled from the prior.

\subsubsection{MCMC sampling options}
The \code{tuning_parameters} control the performance of the MCMC sampler, and in particular the variance of the proposal
distributions. The parameter that scales the Metropolis Hastings proposal for the regression coefficients \code{cprop_beta}, is
tuned to achieve an acceptance ratio in the region of $30\% \sim 40\%$. The baseline hazard parameters for the proposal prior
\code{a_lambda} and \code{b_lambda} are set to 0.01 and \code{alpha} = 0.3. The probability of a birth move is set to the default
value of 0.5.
\begin{Schunk}
\begin{Sinput}
R> tuning_parameters_sim <- list(cprop_beta = 1.35,
+                                cprop_beta_0 = 1.35,
+                                a_lambda = 0.01,
+                                b_lambda = 0.01,
+                                pi_b = 0.5,
+                                alpha = 0.4)
\end{Sinput}
\end{Schunk}
In the next four examples, we explore how to estimate; (i) the baseline hazard using data on the historical controls alone, 
(ii) the baseline hazard and treatment effect using data from the current trial alone (e.g. without borrowing) and 
(iii) the baseline hazard and treatment effect using data from the current trial and from the historical controls
(e.g. with borrowing). We perform the estimation for both marginal and conditional treatment effect estimands. In the context of a marginal treatment estimand, we demonstrate how an adjusted model can be used via G-computation. 

\subsubsection{Current trial / historical controls inference only}
The model can be fitted to the historical controls alone for an estimate of  the posterior baseline hazard with a call similar to 
the one shown below. The first argument, \code{formula}, names the time to event and event indicator variables, \code{tte} and 
\code{event}, wrapped in the \code{Surv} function as is typically done in survival models in \proglang{R}. On the
right hand side we specify either an intercept only model, \code{~1}, or any pre-treatment variables related to
randomisation. These can be continuous or factor variables, or character intended for coercion to factor. The reference level can be set
 in the usual way using an internal function \proglang{R} such as \code{relevel}. The arguments \code{data} and \code{CntlOnly} are  
set to \code{data=DAT_hst},  and \code{CntlOnly=TRUE}. The model \code{hyperparameters} and \code{tuning_parameters}, discussed above, 
are specified. We select 6000 iterations for the sampler after 2000 warmup iterations. 
\begin{Schunk}
\begin{Sinput}
R> fit_sim_hst_cndl <-
+      BayesFBHborrow(formula=Surv(tte, event)~X_01+X_02+X_03,
+                     data = DAT_hst, CntlOnly = TRUE,
+                     hyperparameters = hyperparameters_sim,
+                     tuning_parameters = tuning_parameters_sim,                
+                     warmup_iter = 2000, iter = 6000,
+                     refresh = 2000, verbose = TRUE)
\end{Sinput}
\end{Schunk}
A plot of the baseline hazard can be used to judge the adequacy of the choice of the smoothing parameters, (\code{J_max}, \code{phi}, \code{c_lambda}). Use of
the plot method for the class is described below. 

The model can also be fitted to the current trial alone using a call similar to the above, but with the following changes. The \code{data} argument is set to 
the current trial dataset, \code{DAT_cc}, and the treatment indicator variable is listed first among the regressors on the right side of the formula argument. 
There is no need to set the \code{CntlOnly} argument to \code{FALSE}, as this is its default value. 
\begin{Schunk}
\begin{Sinput}
R> fit_sim_nb_cndl <-
+      BayesFBHborrow(formula=Surv(tte, event)~X_trt+X_01+X_02+X_03,
+                     data = DAT_cc,
+                     hyperparameters = hyperparameters_sim,
+                     tuning_parameters = tuning_parameters_sim,                
+                     warmup_iter = 2000, iter = 6000,
+                     refresh = 2000, verbose = TRUE)
\end{Sinput}
\end{Schunk}

\subsubsection{Inference on current trial with borrowing, with covariates}
To undertake inference on the current trial, whilst borrowing from historical controls, the following changes are made to the
previous call. In addition to specification of the argument \code{data} with data on the current trial, we must now specify the
argument \code{data_hist} with a \code{data.frame} containing data on historical controls. The borrowing model, which determines
the parameterisation of the variance of the baseline hazard prior, is selected using the argument \code{model_choice}. With
\code{model_choice="all"}, the variance of the log baseline hazard is aggregated across all split points. Alternatively, it can be
defined separately for each split point using \code{model_choice="mix"}  (the default). There is also an option for the standard
univariate inverse gamma for each split via \code{model_choice="uni"}, but a ``lump-and-smear'' mixture is recommended.  Here we
use the default option, 
\code{model_choice="mix"}.
\begin{Schunk}
\begin{Sinput}
R> fit_sim_wb_cndl <-
+     BayesFBHborrow(formula=Surv(tte, event)~X_trt+X_01+X_02+X_03,
+                    data = DAT_cc, data_hist = DAT_hst,
+                    model_choice="mix",
+                    hyperparameters = hyperparameters_sim,
+                    tuning_parameters = tuning_parameters_sim,                
+                    warmup_iter = 2000, iter = 6000,
+                    refresh = 2000, verbose = TRUE)
\end{Sinput}
\end{Schunk}

\subsubsection{Output: Conditional Estimand versus Marginal Contrast}
The print method shows two tables: treatment/pretreatment effects and survival probabilities, shown here formatted as booktabs
tables. The treatment/pretreatment effects correspond to the model log hazard ratios for each MCMC sample and are summarised by 
the posterior median together with the $95\%$ credible interval, as shown in Table \ref{tbl:fit_sim_wb_cndl_coef}.
\begin{table}[H]
  \begin{center}
  \caption{Conditional Treatment Effect and Pre-treatment Covariates Effects in model borrowing from historical controls.}
  \label{tbl:fit_sim_wb_cndl_coef}
\begin{tabular}{lrrrr}
\toprule
  & logHR & exp(logHR) & CrIn.L & CrIn.H\\
\midrule
X\_trt & -0.9406 & 0.3904 & -1.2242 & -0.6108\\
X\_01 & -0.2737 & 0.7606 & -0.4273 & -0.1315\\
X\_02 & 0.5477 & 1.7292 & 0.3908 & 0.7257\\
X\_03b & 0.6625 & 1.9397 & 0.4532 & 0.8482\\
X\_03c & -0.4117 & 0.6625 & -0.6546 & -0.1735\\
X\_01\_0 & -0.1059 & 0.8995 & -0.3949 & 0.1779\\
X\_02\_0 & 0.5432 & 1.7216 & 0.3645 & 0.7429\\
X\_03b\_0 & 0.2124 & 1.2367 & -0.0703 & 0.5106\\
X\_03c\_0 & -0.2394 & 0.7871 & -0.5356 & 0.0510\\
\bottomrule
\end{tabular}  \end{center}
\end{table}
Table \ref{tbl:fit_sim_wb_cndl_coef} reveals a statistically significant treatment effect, with the conditional treatment effect indicating a risk
reduction of 61\%, and the credible interval fully on one side of zero.

Displayed survival probabilities are by default conditional per arm survival probabilities per MCMC sample e.g. $S(t_f)$, in the
control arm and $S(t_f)^{\e^{\beta_{\mrm{trt}}}}$ in the intervention arm at $f=25\%, 50\%, 75\%$ and 100\% information time. Here, $t_f$ denotes the time at which the indicated portion, $f$ of total events is reached. These quantities are again summarised by the posterior median together with the $95\%$ credible interval. 

Notice that here ``conditional'' means that each of these estimates represents survival in a hypothetical group of patients with pretreatment covariates equal to their referent values. For continuous variables, these are their means over the dataset supplied in the argument \code{data}. For factor variables, they are also set at their means, but under a hypothetical design balanced across the levels of each factor. These results are shown in Table \ref{tbl:fit_sim_wb_cndl_surv}.
\begin{table}[H]
  \begin{center}
  \caption{Conditional per arm survival probabilities at landmark information fraction in model with borrowing from historical
    controls, simulated data.}
  \label{tbl:fit_sim_wb_cndl_surv}
\begin{tabular}{llrrrr}
\toprule
Arm & InfFrac & Time & Survival & Surv\_L & Surv\_U\\
\midrule
C & 0.25 & 0.5243 & 0.9744 & 0.9650 & 0.9823\\
C & 0.50 & 0.8177 & 0.9362 & 0.9185 & 0.9520\\
C & 0.75 & 1.0378 & 0.8997 & 0.8754 & 0.9211\\
C & 1.00 & 1.4966 & 0.8165 & 0.7699 & 0.8569\\
I & 0.25 & 0.5243 & 0.9899 & 0.9852 & 0.9934\\
I & 0.50 & 0.8177 & 0.9744 & 0.9656 & 0.9814\\
I & 0.75 & 1.0378 & 0.9593 & 0.9474 & 0.9687\\
I & 1.00 & 1.4966 & 0.9236 & 0.8993 & 0.9417\\
\bottomrule
\end{tabular}  \end{center}
\end{table}
  
\subsubsection{G-computation}
When the model is fit to a trial, either with or without borrowing, and pre-treatment covariates are included, then after the 
sampler runs and conditional estimands are derived, G-computation is performed to obtain the marginal per arm survival 
probabilities and marginal treatment effect (MTE). As the MTE is the difference of log-logged marginal per arm survival function 
averaged over pretreatment variables, it is a time-varying quantity and is therefore shown at landmark information fractions in 
Table \ref{tbl:mgnl_estmd}. 

The print method is coerced into displaying marginal contrasts by specifying \code{G_compute=TRUE} either in the original call 
sequence, or in an application of the \code{update} method. The latter is done without any need to rerun the sampler. 
\begin{Schunk}
\begin{Sinput}
R> mgnl_cntrst_sim_wb <- update(fit_sim_wb_cndl, G_compute=TRUE)
\end{Sinput}
\end{Schunk}

\begin{table}[H]
  \begin{center}
  \caption{Marginal treatment effect at landmark information fraction in model with borrowing from historical controls, simulated data.}
  \label{tbl:mgnl_estmd}
\begin{tabular}{lrrrrr}
\toprule
InfFrac & Time & MTE & exp(MTE) & MTE\_L & MTE\_U\\
\midrule
0.25 & 0.5243 & -0.2339 & 0.7914 & -0.3099 & -0.1448\\
0.50 & 0.8177 & -0.2205 & 0.8022 & -0.2913 & -0.1378\\
0.75 & 1.0378 & -0.2143 & 0.8071 & -0.2832 & -0.1345\\
1.00 & 1.4966 & -0.2027 & 0.8166 & -0.2721 & -0.1270\\
\bottomrule
\end{tabular}  \end{center}
\end{table}
Some final comments regarding the calling sequence. Notice when this formula/data interface to the function is
used, variable names are entirely immaterial, with the only unusual requirement that the treatment indicator, when included, must
be listed first, in addition to consistency of variable names among all datasets when there is more than one supplied. This
provision is recommended in most cases. There is also a data only interface which offers additional flexibility, e.g. allows
interactions et cetera. This option is used by omitting the \code{formula} argument, and instead, adhering to the following strict
variable naming conventions. The time to event and event indicator variables must be named \code{tte} and \code{event}
respectively, and variable names must begin with the prefix \code{X_}. Again, a treatment indicator, when used, must be the first
of these so named variables in all datasets. Following this approach, factor variables can still be left as is or as character
with intention to coerce as factor, and the function will recode them to dumy indcator variables by default, but without any
possibility of entering interaction terms. If the user wishes to include interactions then in addition to using the data only
interface, the argument \code{preprocess} must be set to \code{FALSE} and all recoding must be done manually.

For publication purposes, the user can extract the tables shown above for downstream wrapping in a table formatter of their
own choosing. For example, if \code{my_fit} is the result of a call with \code{G_compute=FALSE} then conditional estimand tables
are extracted via the following commands:
\begin{itemize}
  \item[-]{Coefficients Table: \code{coef(my_bfbhb)}.}
  \item[-]{Survival Probabilities Table: \code{summary(my_bfbhb)$surv_summary}.}
\end{itemize}
with marginal contrast versions obtained by wrapping the object first in a call to the update method \code{update(my_bfbhb,
  G_compute=TRUE)}. Note as well that the survival probabilities at landmark information fraction are internally extracted from
the more detailed version which contains corresponding survival probabilities at every time in the fine meshed internal grid of
length \code{max_grid} (default 2000). These are in the component \code{surv_dat} of the returned object, representing
conditional estimand or marginal contrast depending upon the value of \code{G_compute}.

\subsubsection{Output--MCMC Sampling details}
The complete sampler history is recorded and accessible, allowing derivation of any posterior functional. Specifically, the
\code{samples} component of the returned object is a data.frame with \code{iter} rows, containing all scalar-valued components
such as: 
\begin{itemize}
  \item[-]{Model coefficients (for current and historical data likelihood portions)}
  \item[-]{Number and locations of split points ($J$, $s_0, s_1, \ldots, s_J, s_{J+1}$)}
  \item[-]{Sampled hyperparameters, such as ($\mu$, $\sigma^2_\lambda$) from NN-GMRF prior on the baseline hazard values.}
  \item[-]{Piecewise constant baseline hazards ($\lambda_1,\cdots, \lambda_J$ and $\lambda_{0,1}, \cdots, \lambda_{0,J}$), for
    both current and historical data (maximum split points: $J$)}
\end{itemize}
While \code{samples} provides a comprehensive sampler output representation, the most practical form for baseline hazard samples is
their snapped-to-common-grid version (each sample of length \code{max_grid} per sample). This large matrix is written as a binary 
file to facilitate efficient storage, with a unique file name that is generated and stored in the returned object.

If the full snapped-to-common-grid matrix of samples (dimensions: \code{max_grid} $\times$ \code{iter}) for the conditional 
baseline hazard is needed, use the function \code{read_haz_mcmc_smpls} as shown below. 
\begin{Schunk}
\begin{Sinput}
R> surv_dat_full <- read_haz_mcmc_smpls(fit_sim_wb_cndl)
\end{Sinput}
\end{Schunk}
The result is a list with 3 components, \code{cnd_haz_0}, \code{mgnl_haz_0} and \code{mgnl_haz_1}. The first is the conditional baseline 
hazard while the last two are the log-logged arm specific survival functions from which the marginal treatment effect is derived as the 
difference.

\subsubsection{Plot method}
\begin{figure}[H]
  \begin{center}
   \caption{Estimated conditional per arm hazard functions in fit to current trial alone (left pane) and with borrowing from
     historical controls (right pane), simulated data}
   \label{fig:haz-nb-wb}
\includegraphics[width=0.7\textwidth]{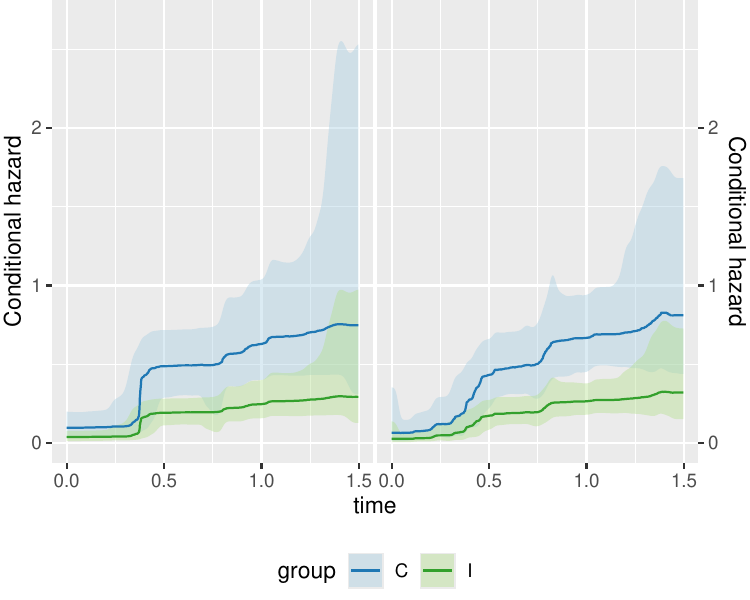}
\end{center}
\end{figure}

The plot method for the \code{BayesFBHborrow} class will create upon the user's choice, a plot of control and intervention arm
hazards or survival functions in the form of a median coupled with 95\% pointwise credible interval ribbons, or density plot of
the treatment effect (conditional or marginal at specified landmark information fraction) or a trace plot of any one of the state
variables in the output component \code{samples}. These will have a differing interpretation depending on whether
\code{G_compute} was set to \code{FALSE} or \code{TRUE} as described above, namely that the relevent quantities are conditional or
marginal estimates from the G-computation.  A plot of the hazard function estimated from historical controls is created by 
\begin{Schunk}
\begin{Sinput}
R> p_cndl_haz_hst <- plot(fit_sim_hst_cndl, type="hazard")
\end{Sinput}
\end{Schunk}
This is used to check the adequacy of the smoothing parameters, (\code{J_max}, \code{phi}, \code{c_lambda}), as mentioned above, and 
as displayed in the Appendix (Figure \ref{fig:cndl_haz_hst_cc}). The next two lines create plots of the per arm conditional hazard 
functions from the fit to the current trial without borrowing, and fit to the current trial with borrowing from historical controls, 
respectively.
\begin{Schunk}
\begin{Sinput}
R> p_cndl_haz_nb <- plot(fit_sim_nb_cndl, type="hazard")
R> p_cndl_haz_wb <- plot(fit_sim_wb_cndl, type="hazard")
\end{Sinput}
\end{Schunk}
The \fct{Combine} function is used to neatly combine two of the plots as shown in Figure \ref{fig:haz-nb-wb}.
\begin{Schunk}
\begin{Sinput}
R> p_cndl_haz_cmb <- Combine(p_cndl_haz_nb, p_cndl_haz_wb)
\end{Sinput}
\end{Schunk}
Corresponding plots of conditional per arm survival curves are obtained by specifying \code{type="survival"} instead.

\subsection{Analysis of German Breast Cancer Study Data}
For the second series of examples, the model is applied to the German Breast Cancer Study Data (gbcsCS) \citep{gbsg_data},
available from the \proglang{R} package \pkg{condSURV} \citep{condsurv}. The original dataset comprises 686 observations across 16
variables, for patients with node-positive breast cancer, recruited between July 1984 and December 1989. Additional patient-level
information in the dataset includes the treatment (hormone), time to recurrence or censoring (rectime), recurrence or censoring
indicator (censrec), tumour grade (grade), and size (size).

We are interested in investigating the effect of treatment with the hormone Tamoxifen on the hazard ratio after adjusting for pretreatment 
variables, and therefore target a conditional estimand. We split the control arm into two using the median time of enrollment, to create 
a historical set and a current set. Thus, there are 193 and 247 patients in the historical and current control groups, and 96 treated patients
who enrolled after the median for the current treated group.

After loading the dataset, we perform some data wrangling. We use \code{diagdateb} to distinguish the current and historical datasets, and define the treatment variable, \code{tamoxifen}, as one less
than the \{1,2\} valued original variable, \code{hormone}. We do the same for the variable \code{menopause} and convert the
original variable, \code{grade} into a factor with levels corresponding to the original ordinal values 1, 2, and 3. We  define a variable \code{hst_flg} to help us artificially create our current trial and historical datasets, based on whether a patient enrolled before or after the median enrollment time. Then, we create a factor variable, \code{group}, which we'll use to construct
Kaplan-Meier plots. Finally, we split the dataset into ``current'' and ``historical'' portions according to the previously mentioned
chronology of enrollement. 
\begin{Schunk}
\begin{Sinput}
R> data(gbcsCS, package="condSURV")
R> gbcs_full <- gbcsCS
R> gbcs_full$diagdateb <- 
+            as.Date(format(as.character(gbcs_full$diagdateb),
+                    format="%d-%m-%Y"), format="%d-%m-%Y")
R> gbcs_full$tamoxifen <- gbcs_full$hormone - 1
R> gbcs_full$menopause <- gbcs_full$menopause - 1
R> gbcs_full$grade <- factor(gbcs_full$grade, levels=as.character(1:3))
R> gbcs_full$hst_flg <- 1*with(gbcs_full, diagdateb < median(diagdateb))
R> grp_lvls <- c("Hist Cntl","Curr Cntl","Curr Trt")
R> gbcs_full$group <-factor(with(gbcs_full,
+                                grp_lvls[tamoxifen + (1 - hst_flg) + 1]),
+                                levels=grp_lvls)
R> gbcs_curr <- gbcs_full[gbcs_full$hst_flg==0,]
R> gbcs_hist <- gbcs_full[with(gbcs_full, (hst_flg==1) & (tamoxifen==0)),]
\end{Sinput}
\end{Schunk}

Here are the first six rows of the current dataset.
\begin{table}[H]
  \begin{center}
  \caption{German Breast Cancer Study Data-- Enrollment after the Median -- Browsing 1st 6 lines}
  \label{tbl:gbcs-show}
\begin{tabular}{rlrrrrrrl}
\toprule
id & diagdateb & rectime & censrec & tamoxifen & menopause & estrg\_recp & size & grade\\
\midrule
17 & 1986-10-14 & 518 & 1 & 0 & 1 & 9 & 20 & 2\\
22 & 1986-10-14 & 1722 & 0 & 1 & 0 & 20 & 21 & 2\\
84 & 1987-10-16 & 535 & 1 & 0 & 0 & 8 & 55 & 3\\
85 & 1987-04-02 & 1653 & 0 & 0 & 1 & 14 & 30 & 2\\
194 & 1987-02-05 & 1182 & 0 & 0 & 0 & 11 & 25 & 2\\
195 & 1987-03-20 & 71 & 0 & 0 & 0 & 41 & 18 & 2\\
\bottomrule
\end{tabular}\end{center}
\end{table}

We take a look at the Kaplan-Meier plots blocked on chronological dataset and treatment group.
\begin{Schunk}
\begin{Sinput}
R> grp_lvls <- c("Hist Cntl","Curr Cntl","Curr Trt")
R> fit_sf <- survfit(Surv(rectime, censrec)~group, data=gbcs_full)
R> KM_dat <- data.frame(time=fit_sf$time, surv=fit_sf$surv)
R> n_strat <- fit_sf$strata
R> KM_dat$group <- c(rep(grp_lvls[1], n_strat[1]), 
+                    rep(grp_lvls[2], n_strat[2]),
+                    rep(grp_lvls[3], n_strat[3]))
R> p_KM <- ggplot(data=KM_dat) + 
+          geom_step(aes(time, surv, group=group, color=group))
\end{Sinput}
\end{Schunk}

\begin{figure}[H]
  \begin{center}
    \caption{Kaplan-Meier Plot for German Breast Cancer Data Stratified on Enrollment before or after median and Intervention Arm}
    \label{fig:gbcsKM}
\includegraphics[width=0.7\textwidth]{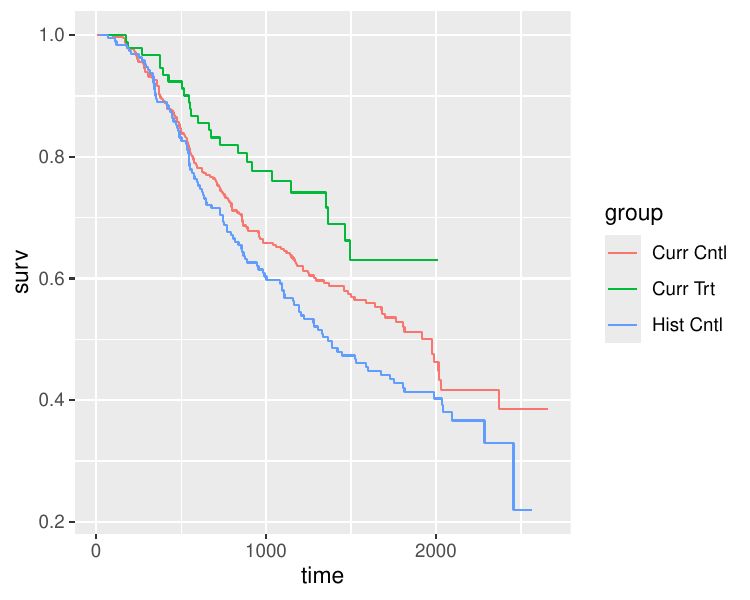}
\end{center}
\end{figure}
The Kaplan-Meier plots (Figure \ref{fig:gbcsKM}) appear to exhibit a drift in the marginal probability of survival over time. This may
be caused by various factors such as improvements in the underlying standard of care or changes in the patient demographics.  This 
motivates the importance of covariate adjusted borrowing, which will account for drift if it is driven by the pretreatment variables.

We adopt a conservative approach to borrowing, through the principled approach to selection of borrowing hyperparameters outlined in
\cite{Scott2024}, in order to address the possible drift between the control groups. 
We set $a_\tau=c_\tau=1$, $b_\tau=0.001$ and $d_\tau=5$ (as in our previous example) and define a limit of tolerable difference
for the current and historical standardised log baseline hazard of approximately $\xi=0.18$, which leads to a prior weight of 0.5
(the borrowing profile is in Figure \ref{fig:qpost_1}).

\begin{Schunk}
\begin{Sinput}
R> xifinder(b_tau = 0.001, d_tau = 5, p_0 = 0.5)
\end{Sinput}
\begin{Soutput}
[1] 0.1797401
\end{Soutput}
\end{Schunk}

\begin{Schunk}
\begin{Sinput}
R> hyperparameters_gbcs <- list(beta_prior = 10^2,
   +                              beta_0_prior = 10^2,
   +                              a_tau = 1,
   +                              b_tau = 0.001,
   +                              c_tau = 1,
   +                              d_tau = 5,
   +                              p_0 = 0.5,
   +                              a_sigma = 1,
   +                              b_sigma = 1,
   +                              clam_smooth = 0.8,
   +                              phi = 3,
   +                              Jmax = 5)
\end{Sinput}
\end{Schunk}

We specify the variance on our vague prior on the regression coefficients of the historical study with \code{beta_0_prior}=$10^2$,
and both the treatment effect and regression coefficients in the control group with \code{beta_prior}=$10^2$. The MCMC sampler proposal scalars,
\code{cprop_beta} and \code{cprop_beta_0}, are tuned by running subsequent models, with much fewer total iterations, so that we reach an
acceptance ratio of $30\% \sim 40\%$. For example, we specify the following tuning parameters; 
\begin{Schunk}
\begin{Sinput}
R> tuning_parameters_gbcs <- list(cprop_beta = 1.17,
+                                 cprop_beta_0 = 1.21,
+                                 a_lambda = 0.5,
+                                 b_lambda = 0.5,
+                                 pi_b = 0.5,
+                                 alpha = 0.4)
\end{Sinput}
\end{Schunk}
and conduct a test run with 500 iterations to check the acceptance ratios,
\begin{Schunk}
\begin{Sinput}
R> test <- 
+      BayesFBHborrow(Surv(rectime, censrec)~tamoxifen + menopause + 
+                                             size + grade,
+                     data=gbcs_curr, data_hist=gbcs_hist,
+                     borrow=TRUE, model_choice="mix",
+                     tuning_parameters=tuning_parameters_gbcs,
+                     hyperparameters=hyperparameters_gbcs,
+                     iter=500, warmup_iter=50, refresh=0)
\end{Sinput}
\end{Schunk}
leading to acceptable acceptance ratios of \code{beta_acc_ratio}=0.316 and \code{beta_0_acc_ratio}=0.366, respectively. 

Our first model fit, for illustration purposes, is without borrowing. Our current dataset is \code{gbcs_curr} and the treatment variable is \code{tamoxifen}. We adjust for the pretreatment variables \code{menopase} (0/1) \code{size} (tumour size, continuous) and
\code{grade} (histologic grade of tumour, factor). 
\begin{Schunk}
\begin{Sinput}
R> fit_gbcs_nb_cndl <- 
+      BayesFBHborrow(Surv(rectime, censrec)~tamoxifen + menopause + 
+                                            size + grade,
+                     data=gbcs_curr,
+                     tuning_parameters=tuning_parameters_gbcs,
+                     hyperparameters=hyperparameters_gbcs,
+                     iter=6000, warmup_iter=2000, refresh=2000,
+                     verbose=TRUE)
\end{Sinput}
\end{Schunk}

Next, we fit the model with borrowing from historical controls. 
\begin{Schunk}
\begin{Sinput}
R> fit_gbcs_wb_cndl <- 
+      BayesFBHborrow(Surv(rectime, censrec)~tamoxifen + menopause + 
+                                             size + grade,
+                     data=gbcs_curr, data_hist=gbcs_hist,
+                     borrow=TRUE, model_choice="mix",
+                     tuning_parameters=tuning_parameters_gbcs,
+                     hyperparameters=hyperparameters_gbcs,
+                     iter=6000, warmup_iter=2000, refresh=2000,
+                     verbose=TRUE,
+                     max_grid=2000,
+                     standardise=TRUE)
\end{Sinput}
\end{Schunk}
resulting in the treatment effect estimate in Table~\ref{tbl:fit_gbcs_wb_cndl_coef}. The convergence of the sampler is checked by inspecting the trace plot of the parameters, such as the conditional treatment
effect \code{tamoxifen} in Figure \ref{fig:trace}.
\begin{Schunk}
\begin{Sinput}
R> p_gbcs_wb_trace <- plot(fit_gbcs_wb_cndl, type="trace", col="tamoxifen")
\end{Sinput}
\end{Schunk}

\begin{table}[H]
  \begin{center}
  \caption{Conditional Treatment Effect and Pre-treatment Covariates Effects in model borrowing from historical controls.}
  \label{tbl:fit_gbcs_wb_cndl_coef}
\begin{tabular}{lrrrr}
\toprule
  & logHR & exp(logHR) & CrIn.L & CrIn.H\\
\midrule
tamoxifen & -0.4564 & 0.6336 & -0.7871 & -0.1379\\
menopause & -0.0630 & 0.9389 & -0.3171 & 0.1686\\
size & 0.0110 & 1.0111 & 0.0033 & 0.0191\\
grade2 & 0.2165 & 1.2417 & 0.0151 & 0.4274\\
grade3 & 0.2765 & 1.3186 & 0.0582 & 0.5162\\
menopause\_0 & 0.3438 & 1.4103 & 0.0855 & 0.6021\\
size\_0 & 0.0194 & 1.0196 & 0.0108 & 0.0279\\
grade2\_0 & 0.3175 & 1.3737 & 0.0705 & 0.5854\\
grade3\_0 & 0.4717 & 1.6026 & 0.2282 & 0.7681\\
\bottomrule
\end{tabular}  \end{center}
\end{table}
The conditional treatment effect of tamoxifen is a reduction in recurrence of
36.6\%, with credible interval ranging from
12.9\% to 54.5\%.

As previously mentioned one can look at the predictive curves with calls to the \fct{plot} methods.
\begin{Schunk}
\begin{Sinput}
R> p_gbcs_cnd_nb_haz <- plot(fit_gbcs_nb_cndl, 
+                            type="hazard", ylim=c(0,0.0010))
R> p_gbcs_cnd_wb_haz <- plot(fit_gbcs_wb_cndl, 
+                            type="hazard", ylim=c(0,0.0010))
R> p_gbcs_cnd_haz_cmb <- Combine(p_gbcs_cnd_nb_haz, p_gbcs_cnd_wb_haz)
R> p_gbcs_cnd_nb_TrtEff <- plot(fit_gbcs_nb_cndl, type="TrtEff")
R> p_gbcs_cnd_wb_TrtEff <- plot(fit_gbcs_wb_cndl, type="TrtEff")
R> p_gbcs_cnd_TrtEff_cmb <- Combine(p_gbcs_cnd_nb_TrtEff, 
+                                   p_gbcs_cnd_wb_TrtEff, lgnd=c("NB","WB"))
\end{Sinput}
\end{Schunk}
\begin{figure}[H]
  \begin{center}
  \caption{Conditional Estimate of baseline hazard function from model fit to current trial alone and fit with borrowing from
    historical controls for the German Breast Cancer Study}
\includegraphics[width=0.7\textwidth]{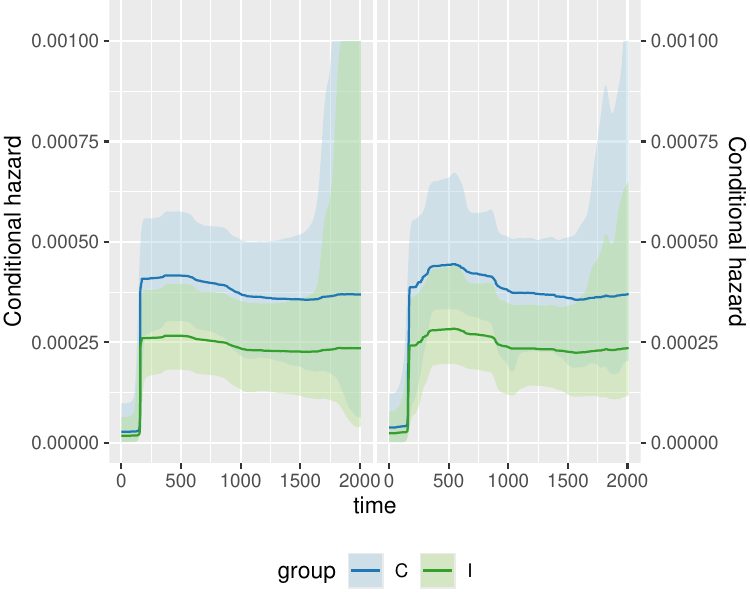}
\end{center}
\end{figure}

\begin{figure}[H]
  \begin{center}
  \caption{Posterior density of conditional treatment effect from model fit to current trial alone and fit with borrowing from
    historical controls for the  German Breast Cancer Study} 
\includegraphics[width=0.7\textwidth]{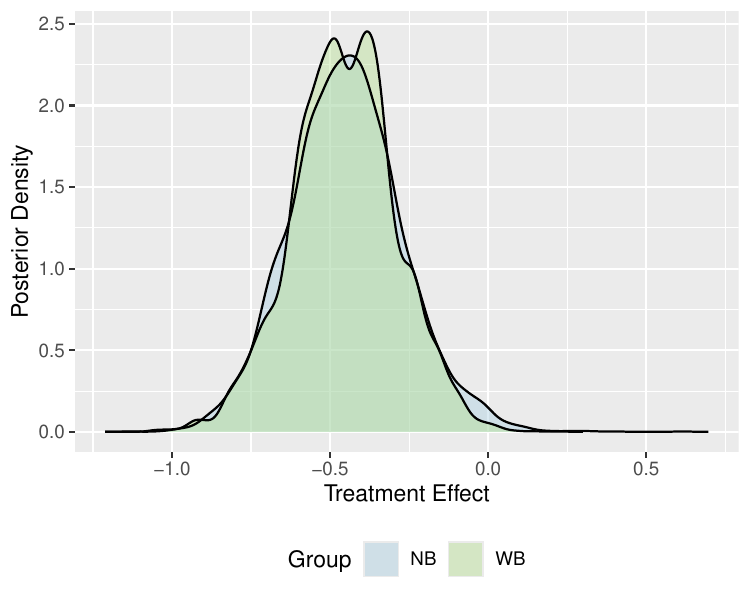}
\end{center}
\end{figure}

\section{Summary} \label{sec:summary}

The \pkg{BayesFBHborrow} package enables Bayesian borrowing within a joint hierarchical model from historical control data in a
time-to-event setting. By integrating over the uncertainty of the location and number of split points in our ensemble method and
incorporating smoothing priors, we induce a flexible baseline hazard in our joint semiparametric
model. This leads to improved borrowing characteristics compared to standard dynamic approaches, which constrain the shape of the
baseline hazard. The borrowing is controlled by a commensurability parameter $\tau_j$, where a prior mixture ensures that the
borrowing is more robust to prior data-data conflict. The model allows for different types of estimands through the addition of
prognostic covariates, which may also improve the precision of the estimate.

Possible extensions of the \pkg{BayesFBHborrow} package include the implementation of borrowing across multiple historical
datasets and a more varied joint prior structure to capture the between trial heterogeneity. With a similar structure to our
commensurate prior (\ref{eq:mix}), \cite{Neuenschwander2016}, proposed an extension to allow for
non-exchangeability. Alternatively, to account for departures from the exchangeability assumption, a Dirichlet process mixture
prior \citep{Escobar1995} could be used to create a data-dependent clustering mechanism. This approach has been adopted by
\cite{Hupf2021} in the context of a binary end point and in a time-to-event setting \citep{Bi2023}. As the number of historical
trials is usually small, this type of clustering can also be achieved using a RJMCMC approach.

Currently there is an option to run an analysis with and without borrowing to quantify the amount of information within the prior
in terms of an historical ESS. This retrospective measure can only be performed once the current data has
been realised. When there is conjugacy, the Fisher information can be used to quantify the prior information centred at the
likelihood \cite{Neuenschwander2020}. Further research is required to be able to quantify the prospective information in the prior for more complicated models.

\section*{Disclosure}
Darren Scott and Grant Izmirlian are employed full-time and possess equity holdings in AstraZeneca.

\bibliography{2026-06-11-Scott-Izmirlian-JSS-joint-2}

@article{oncology,
author = {Su, Liwen and Chen, Xin and Zhang, Jingyi and Yan, Fangrong},
title = {Comparative Study of Bayesian Information Borrowing Methods in Oncology Clinical Trials},
journal = {JCO Precision Oncology},
volume = {6},
number = {6},
pages = {e2100394},
year = {2022},
doi = {10.1200/PO.21.00394},
note ={PMID: 35263169},
URL = {https://doi.org/10.1200/PO.21.00394}
}

@article{cardiovascular,
  title     = "A Practical Bayesian Adaptive Design Incorporating Data from Historical Controls",
  author    = "Psioda, Matthew A and Soukup, Mat and Ibrahim, Joseph G",
  journal   = "Stat. Med.",
  publisher = "Wiley",
  volume    =  37,
  number    =  27,
  pages     = "4054--4070",
  month     =  nov,
  year      =  2018,
  copyright = "http://onlinelibrary.wiley.com/termsAndConditions\#vor",
  language  = "en"
}

@article{rbest,
    title = {Applying Meta-Analytic-Predictive Priors with the {R}
      {B}ayesian Evidence Synthesis Tools},
    author = {Sebastian Weber and Yue Li and John W. Seaman and
      Tomoyuki Kakizume and Heinz Schmidli},
    journal = {Journal of Statistical Software},
    year = {2021},
    volume = {100},
    number = {19},
    pages = {1--32},
    doi = {10.18637/jss.v100.i19}
  }

@manual{psborrow,
    title = {psborrow2: An R Package for Bayesian Dynamic Borrowing Simulation Study and
Analysis},
    author = {Isaac Gravestock},
    year = {2023},
    note = {R package version 0.0.2.0}
  }

@manual{R,
    title = {R: A Language and Environment for Statistical Computing},
    author = {{R Core Team}},
    organization = {R Foundation for Statistical Computing},
    address = {Vienna, Austria},
    year = {2021},
    url = {https://www.R-project.org/}
  }

@article{lewis,
author = {Connor Jo Lewis, Somnath Sarkar, Jiawen Zhu and Bradley P. Carlin},
title = {Borrowing from historical Control Data in Cancer Drug Development: A Cautionary Tale and Practical Guidelines},
journal = {Statistics in Biopharmaceutical Research},
volume = {11},
number = {1},
pages = {67-78},
year = {2019},
publisher = {Taylor & Francis},
doi = {10.1080/19466315.2018.1497533},
note ={PMID: 31435458},
URL = { https://doi.org/10.1080/19466315.2018.1497533}
}

@article{Bi2023,
   author = {Dehua Bi and Meizi Liu and Jianchang Lin and Rachael Liu},
   doi = {10.1002/sim.9936},
   issn = {10970258},
   journal = {Statistics in Medicine},
   publisher = {John Wiley and Sons Ltd},
   title = {BEATS: Bayesian Hybrid Design with Flexible Sample Size Adaptation for Time-to-Event Endpoints},
   year = {2023},
}

@article{Roychoudhury2020,
   author = {Satrajit Roychoudhury and Beat Neuenschwander},
   doi = {10.1002/sim.8456},
   issn = {10970258},
   issue = {7},
   journal = {Statistics in Medicine},
   month = {3},
   pages = {984-995},
   pmid = {31985077},
   publisher = {John Wiley and Sons Ltd},
   title = {Bayesian Leveraging of Historical Control Data for a Clinical Trial with Time-to-Event Endpoint},
   volume = {39},
   year = {2020},
}

@article{Han2017,
   author = {Baoguang Han and Jia Zhan and Z. John Zhong and Dawei Liu and Stacy Lindborg},
   doi = {10.1002/pst.1815},
   issn = {15391612},
   issue = {4},
   journal = {Pharmaceutical Statistics},
   month = {7},
   pages = {296-308},
   pmid = {28560815},
   publisher = {John Wiley and Sons Ltd},
   title = {Covariate-Adjusted Borrowing of Historical Control Data in Randomized Clinical Trials},
   volume = {16},
   year = {2017},
}

@article{hobbs,
author = {Brian P Hobbs and Bradley P Carlin and Daniel J Sargent},
title ={Adaptive Adjustment of the Randomization Ratio Using Historical Control Data},
journal = {Clinical Trials},
volume = {10},
number = {3},
pages = {430-440},
year = {2013},
doi = {10.1177/1740774513483934},
note ={PMID: 23690095},
URL = {https://doi.org/10.1177/1740774513483934},
eprint = {https://doi.org/10.1177/1740774513483934}
}

@article{green,
 ISSN = {00063444},
 URL = {http://www.jstor.org/stable/2337340},
 author = {Peter J. Green},
 journal = {Biometrika},
 number = {4},
 pages = {711--732},
 publisher = {[Oxford University Press, Biometrika Trust]},
 title = {Reversible Jump Markov Chain Monte Carlo Computation and Bayesian Model Determination},
 urldate = {2023-12-11},
 volume = {82},
 year = {1995}
}

@article{pocock,
title = {The Combination of Randomized and Historical Controls in Clinical Trials},
journal = {Journal of Chronic Diseases},
volume = {29},
number = {3},
pages = {175-188},
year = {1976},
issn = {0021-9681},
doi = {https://doi.org/10.1016/0021-9681(76)90044-8},
url = {https://www.sciencedirect.com/science/article/pii/0021968176900448},
author = {Stuart J. Pocock}
}

@ARTICLE{gibbs,
  author={Geman, Stuart and Geman, Donald},
  journal={IEEE Transactions on Pattern Analysis and Machine Intelligence}, 
  title={Stochastic Relaxation, Gibbs Distributions, and the Bayesian Restoration of Images}, 
  year={1984},
  volume={PAMI-6},
  number={6},
  pages={721-741},
  doi={10.1109/TPAMI.1984.4767596}}

@article{MH,
    author = {Hastings, W. K.},
    title = {Monte Carlo Sampling Methods Using Markov Chains and Their Applications},
    journal = {Biometrika},
    volume = {57},
    number = {1},
    pages = {97-109},
    year = {1970},
    month = {04},
    issn = {0006-3444},
    doi = {10.1093/biomet/57.1.97},
    url = {https://doi.org/10.1093/biomet/57.1.97},
    eprint = {https://academic.oup.com/biomet/article-pdf/57/1/97/23940249/57-1-97.pdf},
}

@Manual{condsurv,
    title = {condSURV: Estimation of the Conditional Survival Function for Ordered
Multivariate Failure Time Data},
    author = {Luis Meira-Machado and Marta Sestelo},
    year = {2023},
    note = {R package version 2.0.4},
    url = {https://CRAN.R-project.org/package=condSURV},
  }

@article{gbsg_data,
 ISSN = {09641998, 1467985X},
 URL = {http://www.jstor.org/stable/2680468},
 author = {W. Sauerbrei and P. Royston},
 journal = {Journal of the Royal Statistical Society. Series A (Statistics in Society)},
 number = {1},
 pages = {71--94},
 publisher = {[Wiley, Royal Statistical Society]},
 title = {Building Multivariable Prognostic and Diagnostic Models: Transformation of the Predictors by Using Fractional Polynomials},
 urldate = {2023-12-11},
 volume = {162},
 year = {1999}
}

@article{Scott2024,
     author={Darren A. V. Scott and Alex Lewin},
     title={Borrowing from Historical Control Data in a Bayesian Time-to-Event Model with Flexible Baseline Hazard Function}, 
     journal = {Biostatistics},
     volume = {27},
     number = {1},
     year = {2026},
     pages = { 1-19},
     language = {en},
     url = {https://doi.org/10.1093/biostatistics/kxag006}
}

@Manual{BayesFBHborrow,
    title = {BayesFBHborrow: Bayesian Dynamic Borrowing with Flexible Baseline Hazard
Function},
    author = {Darren Scott and Sophia Axillus},
    year = {2024},
    note = {R package version 2.0.1},
    url = {https://CRAN.R-project.org/package=BayesFBHborrow},
  }

@article{Neuenschwander2016,
   author = {Beat Neuenschwander and Simon Wandel and Satrajit Roychoudhury and Stuart Bailey},
   doi = {10.1002/pst.1730},
   issn = {15391612},
   issue = {2},
   journal = {Pharmaceutical Statistics},
   month = {3},
   pages = {123-134},
   pmid = {26685103},
   publisher = {John Wiley and Sons Ltd},
   title = {Robust exchangeability designs for early phase clinical trials with multiple strata},
   volume = {15},
   year = {2016},
}

@article{Escobar1995,
   author = {Michael D. Escobar and Mike West},
   doi = {10.1080/01621459.1995.10476550},
   issn = {1537274X},
   issue = {430},
   journal = {Journal of the American Statistical Association},
   pages = {577-588},
   title = {Bayesian density estimation and inference using mixtures},
   volume = {90},
   year = {1995},
}

@article{Hupf2021,
   author = {Bradley Hupf and Veronica Bunn and Jianchang Lin and Cheng Dong},
   doi = {10.1002/sim.8970},
   issn = {10970258},
   issue = {14},
   journal = {Statistics in Medicine},
   pages = {3385-3399},
   pmid = {33851441},
   title = {Bayesian semiparametric meta-analytic-predictive prior for historical control borrowing in clinical trials},
   volume = {40},
   year = {2021},
}

@article{Neuenschwander2020,
   author = {Beat Neuenschwander and Sebastian Weber and Heinz Schmidli and Anthony O'Hagan},
   doi = {10.1111/biom.13252},
   issn = {15410420},
   issue = {2},
   journal = {Biometrics},
   month = {6},
   pages = {578-587},
   pmid = {32142163},
   publisher = {John Wiley and Sons Inc},
   title = {Predictively consistent prior effective sample sizes},
   volume = {76},
   year = {2020},
}

@article{Willard2024,
   author = {James Willard and Shirin Golchi and Erica E.M. Moodie},
   doi = {10.1177/09622802241227957},
   issn = {14770334},
   issue = {3},
   journal = {Statistical Methods in Medical Research},
   month = {3},
   pages = {480-497},
   pmid = {38327082},
   publisher = {SAGE Publications Ltd},
   title = {Covariate adjustment in Bayesian adaptive randomized controlled trials},
   volume = {33},
   year = {2024},
}

@article{Keil2018,
   author = {Alexander P. Keil and Eric J. Daza and Stephanie M. Engel and Jessie P. Buckley and Jessie K. Edwards},
   doi = {10.1177/0962280217694665},
   issn = {14770334},
   issue = {10},
   journal = {Statistical Methods in Medical Research},
   month = {10},
   pages = {3183-3204},
   pmid = {29298607},
   publisher = {SAGE Publications Ltd},
   title = {A Bayesian approach to the g-formula},
   volume = {27},
   year = {2018},
}

@article{Rubin1981,
   author = {Donald B. Rubin},
   doi = {10.1214/aos/1176345338},
   issn = {0090-5364},
   issue = {1},
   journal = {The Annals of Statistics},
   month = {1},
   title = {The Bayesian bootstrap},
   volume = {9},
   year = {1981}
}

@article{Stensrud2017,
   author = {Mats Julius Stensrud and Morten Valberg and Kjetil Røysland and Odd O. Aalen},
   doi = {10.1097/EDE.0000000000000621},
   issn = {15315487},
   issue = {3},
   journal = {Epidemiology},
   month = {5},
   pages = {379-386},
   pmid = {28244888},
   publisher = {Lippincott Williams and Wilkins},
   title = {Exploring selection bias by causal frailty models the magnitude matters},
   volume = {28},
   year = {2017}
}

@article{Hernn2010,
   author = {Miguel A. Hernán},
   doi = {10.1097/EDE.0b013e3181c1ea43},
   issn = {10443983},
   issue = {1},
   journal = {Epidemiology},
   month = {1},
   pages = {13-15},
   pmid = {20010207},
   title = {The hazards of hazard ratios},
   volume = {21},
   year = {2010}
}

@article{Fay2024,
   author = {Michael P Fay and Fan Li},
   doi = {10.1177/17407745241243308},
   issn = {1740-7745},
   issue = {5},
   journal = {Clinical Trials},
   month = {10},
   pages = {623-635},
   title = {Causal interpretation of the hazard ratio in randomized clinical trials},
   volume = {21},
   url = {https://journals.sagepub.com/doi/10.1177/17407745241243308},
   year = {2024}
}

@article{Daniel2021,
   author = {Rhian Daniel and Jingjing Zhang and Daniel Farewell},
   doi = {10.1002/bimj.201900297},
   issn = {15214036},
   issue = {3},
   journal = {Biometrical Journal},
   month = {3},
   pages = {528-557},
   pmid = {33314251},
   publisher = {John Wiley and Sons Inc},
   title = {Making apples from oranges: Comparing noncollapsible effect estimators and their standard errors after adjustment for different covariate sets},
   volume = {63},
   year = {2021}
}

@techReport{FDA2026,
   author = {FDA},
   title = {Use of Bayesian methodology in clinical trials of drug and biological products},
   year = {2026}
}

@article{Aalen2015,
   author = {Odd O. Aalen and Richard J. Cook and Kjetil Røysland},
   doi = {10.1007/s10985-015-9335-y},
   issn = {1380-7870},
   issue = {4},
   journal = {Lifetime Data Analysis},
   month = {10},
   pages = {579-593},
   title = {Does Cox analysis of a randomized survival study yield a causal treatment effect?},
   volume = {21},
   year = {2015}
}

@article{Robins1986,
   author = {James Robins},
   doi = {10.1016/0270-0255(86)90088-6},
   issn = {02700255},
   issue = {9-12},
   journal = {Mathematical Modelling},
   pages = {1393-1512},
   title = {A new approach to causal inference in mortality studies with a sustained exposure period—application to control of the healthy worker survivor effect},
   volume = {7},
   year = {1986}
}

@article{Besag1995,
   author = {Julian Besag and Charles Kooperberg},
   isbn = {201401:54:06},
   issue = {4},
   journal = {Biometrika},
   pages = {733-779},
   title = {On conditional and intrinsic autoregressions},
   volume = {82},
   year = {1995},
}

@article{Gogtay2021,
   author = {Nithya Jaideep Gogtay and Priya Ranganathan and Rakesh Aggarwal},
   doi = {10.4103/picr.picr_384_20},
   issn = {22295488},
   issue = {2},
   journal = {Perspectives in Clinical Research},
   month = {4},
   pages = {106-112},
   publisher = {Wolters Kluwer Medknow Publications},
   title = {Understanding estimands},
   volume = {12},
   year = {2021}
}

\newpage

\begin{appendix}

\section{G-computation procedure} \label{app:g-computation}

The following G-computation procedure, via a Bayesian bootstrap, is performed in the \pkg{BayesFBHborrow} package to obtain the
marginal log-hazard ratio treatment estimate.  After sampling $m=1,...,M$ Monte Carlo samples from the joint posterior
distribution of the model parameters $\boldsymbol{\theta}|\boldsymbol{D}, \boldsymbol{D}_0$, two copies of the sampled
covariate data $\boldsymbol{X}$ are created, where $Z_i=z$ defines the treatment allocation for all $i = 1, ...,n$. For each
$\boldsymbol{\theta}_s$ and treatment allocation we calculate the posterior marginal hazard;
\begin{enumerate}
  \item For each $i=1,...,n$ the conditional survival probabilities are calculated at time $t$ corresponding to the time
     from the start of the trial to the current analysis.
  \item Sample $\boldsymbol{\pi}_m = (\pi_{m1},....,\pi_{mn})$ from the posterior  distribution $\text{Dirichlet}(1_n)$ for each
    retained iteration of the MCMC sampler $m$. 
  \item Average over these $n$ values to marginalise with respect to the observed covariates,
    \begin{equation*}
      S(t|Z=z,\boldsymbol{\theta}_m) \approx \sum_{i=1}^{n} \pi_{mi}\exp(-H_0(t))^{\exp\{ \vec{x}'_{i}\boldsymbol{\beta}_m +z\varphi_m\}}
    \end{equation*}
    where $i$ in $\vec{x}'_{i}$ identifies the covariate row. This is on the survival probability not the hazard (as the hazard is not a
    probability density but a conditional rate). 
  \item Apply a $\log(-\log(\cdot))$ transformation to yield a single sample from the posterior distribution of $\log(h(t|Z=z))$.
\end{enumerate}
$M$ samples of the posterior distribution of the log-hazard ratio $\gamma(\theta)$ at time $t$ are obtained from
\begin{equation*}
  \gamma(\theta)_m = \log(-\log(S(t|Z=1,\boldsymbol{\theta}_m))- \log(-\log(S(t|Z=0,\boldsymbol{\theta}_m)).
\end{equation*}
Posterior summaries can be easily created for the hazard ratio at any time point. Alternatively a posterior hazard ratio marginalised over
time can be also estimated.

\section{Borrowing profiles} \label{app:borrowing_profiles}
Following Section \ref{sec:hyper_choice}, the choice of the prior weight can be aided by using the borrowing profiles
(\ref{eq:bb_profile}). This is a series of curves mapping the posterior weight as a function of the difference between the
standardised log baseline hazards of the control and historical data, for various prior weights. They illustrate how sensitive the
borrowing is to changes in these differences. As the data is standardised in the algorithm,  we choose our commensurate prior
hyperparameters as $a_\tau=c_\tau=1$, $b_\tau = 0.001$ and $d_\tau=5$. Our choice of 0.8 for the prior weight $p_0$ in the
simulated data application, means our tolerable difference between the current and historical standardised log hazards of 0.29 (or
a hazard ratio of 1.34). Beyond this value, the posterior weight is below 0.5 and smear mixture element is more likely. To be less
conservative with borrowing, we can increase the prior weight. For our applied data example, we choose a tolerable difference
between the current and historical standardised log hazards of 0.18 (or a hazard ratio of 1.20). This yields a prior weight, given
our hyperparameter selection of $p_0=0.5$. Both choices our illustrated, on the set of borrowing profiles (for our
hyperparameters).

\begin{figure}[H]
\centering
\includegraphics{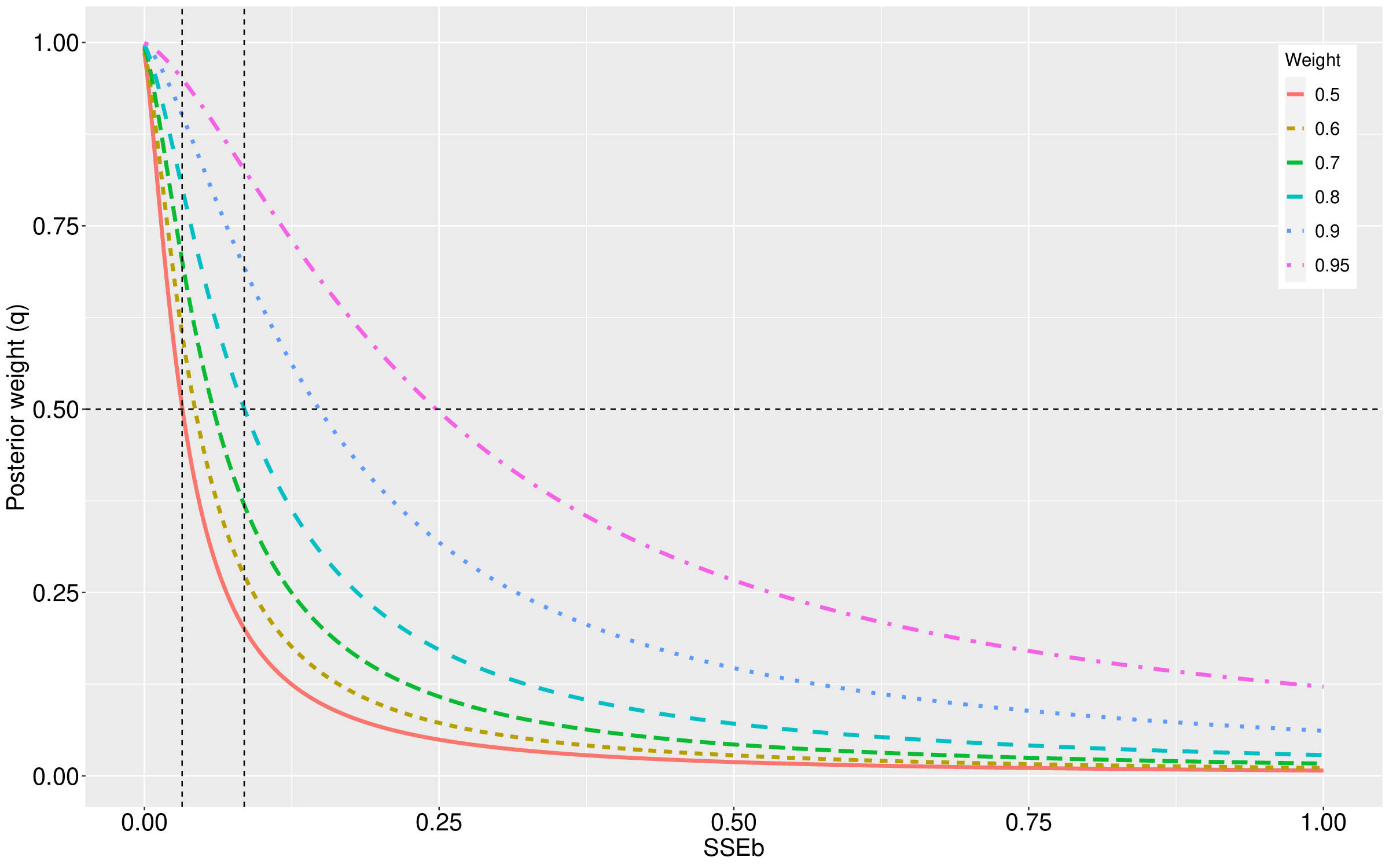}
\caption{\label{fig:qpost_1} Borrowing profiles (posterior weight as a function of the sum of squared error (SSEb) of the
  standardised log baseline hazards) from $b_\tau$ = 0.001 and $d_\tau$ = 5, for  various prior weights ($p_0$). From left to
  right, the first vertical dashed line at $\xi=0.18^2$ represents our threshold of tolerable difference between the log hazards
  (SSEb, or $\xi^2$), leading to a prior weight of $p_0=0.5$. The second vertical dashed line of $\xi= 0.29^2$ is the threshold of
  tolerable difference from a prior weight of $p_0=0.8$. The horizontal dashed line shows where the posterior weight begins to
  fall below the tipping point of 0.5.} 
\end{figure}

\section{Additional Plots } \label{app:add_plots}
\begin{figure}[H]
  \begin{center}
   \caption{Estimated hazard from historical controls (left pane) and per arm hazard from current trial, no borrowing (right pane), simulated data}
     \label{fig:cndl_haz_hst_cc}
\includegraphics[width=0.7\textwidth]{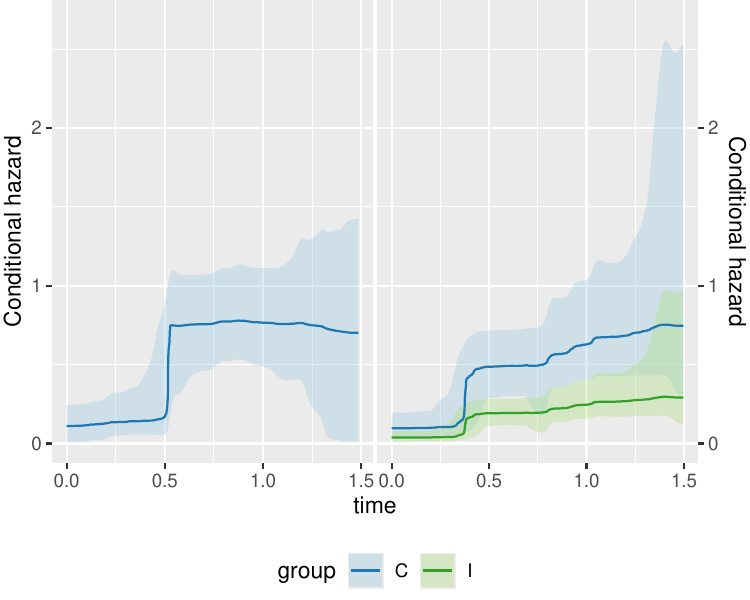}
\end{center}
\end{figure}

\begin{figure}[H]
  \begin{center}
  \caption{Trace plot for MCMC samples of the Conditional Treatment Effect of Tamoxifen, in model fit to current trial with
    borrowing on historical controls, German Breast Cancer Study}
  \label{fig:trace}
\includegraphics[width=0.7\textwidth]{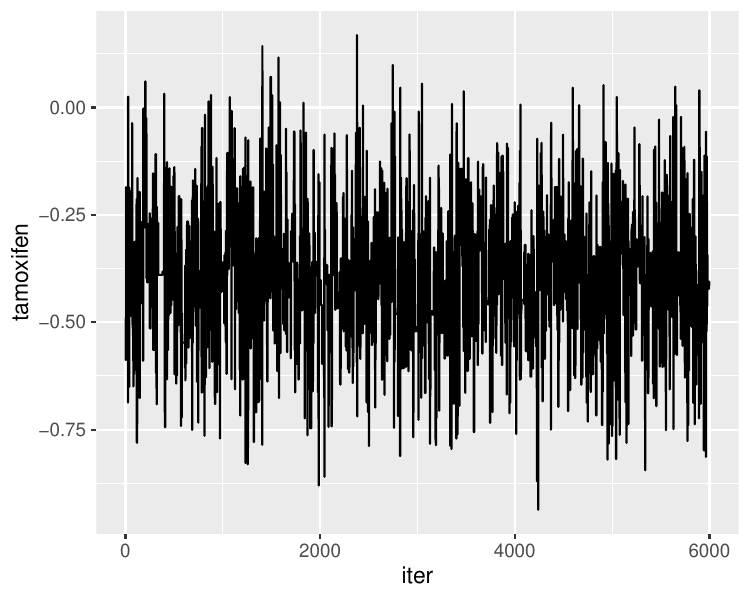}
 \end{center}
\end{figure}

\end{appendix}

\end{document}